\documentclass[aps,prc,preprint,tightenlines,ece,amsmath,groupedaddress]{revtex4} 
\usepackage{epsf}
\usepackage{graphicx}
\usepackage{graphics}
\usepackage{color}
\def\beg{\begin{equation}}
\def\ee{\end{equation}}

\usepackage{longtable}
\begin{document}
\markboth{Raghunath Sahoo} {Relativistic Kinematics \hspace{8cm}   Raghunath Sahoo}
\def\prl{{\em Phys. Rev. Lett. }}
\def\prc{{\em Phys. Rev. C }}
\def\jap{{\em J. Appl. Phys. }}
\def\ajp{{\em Am. J. Phys. }}
\def\nima{{\em Nucl. Instr. and Meth. Phys. A }}
\def\npa{{\em Nucl. Phys. A }}
\def\npb{{\em Nucl. Phys. B }}
\def\epjc{{\em Eur. Phys. J. C }}
\def\plb{{\em Phys. Lett. B }}
\def\mpla{{\em Mod. Phys. Lett. A }}
\def\pr{{\em Phys. Rep. }}
\def\prv{{\em Phys. Rev. }}
\def\zpc{{\em Z. Phys. C }}
\def\zpa{{\em Z. Phys. A }}
\def\yf{{\em Yad. Fiz }}
\def\ppnp{{\em Prog. Part. Nucl. Phys. }}
\def\jpg{{\em J. Phys. G }}
\def\cpc{{\em Comput. Phys. Commun.}}
\def\app{{\em Acta Physica Pol. B }}
\def\aip{{\em AIP Conf. Proc. }}
\def\jhep{{\em J. High Energy Phy. }}
\def\ijmpa{{\em Int. J. Mod. Phys. A }}
\def\anyas{{\em Annals N.Y. Acad.Sci  }}
\def\ppnp{{\em Prog. Part. Nucl. Phys. }}

% Use the \preprint command to place your local institutional report
% number in the upper righthand corner of the title page in preprint mode.
% Multiple \preprint commands are allowed.
% Use the 'preprintnumbers' class option to override journal defaults
% to display numbers if necessary
%\preprint{}
%Title of paper
%\title{Constituent Quarks and Enhancement of Multi-strange Baryons in Heavy-Ion Collisions}
\title{Relativistic Kinematics}
% repeat the \author .. \affiliation  etc. as needed
% \email, \thanks, \homepage, \altaffiliation all apply to the current
% author. Explanatory text should go in the []'s, actual e-mail
% address or url should go in the {}'s for \email and \homepage.
% Please use the appropriate macro foreach each type of information

% \affiliation command applies to all authors since the last
% \affiliation command. The \affiliation command should follow the
% other information
% \affiliation can be followed by \email, \homepage, \thanks as well.

\author{Raghunath Sahoo\footnote{Email: Raghunath.Sahoo@cern.ch,  (Lecture delivered in $IX$ SERC School on Experimental High Energy Physics, IIT Madras, India,12-14 Dec. 2013)}} 

%\email[]{Raghunath.Sahoo@pd.infn.it}

%\homepage[]{Your web page}
%\thanks{}
%\altaffiliation{}
%\affiliation{{$^1$}Dipartimento di Fisica dell'Universit$\grave{a}$
%and Sezione INFN di Padova, Italy}

%\affiliation{Indian Institute of Technology Indore, Indore, India-452017} // original
\address{Indian Institute of Technology Indore, India-452020}

%Collaboration name if desired (requires use of superscriptaddress
%option in \documentclass). \noaffiliation is required (may also be
%used with the \author command).
%\collaboration can be followed by \email, \homepage, \thanks as well.
%\collaboration{}
%\noaffiliation

%\date{\today} % uncomment if you want to put the date

%\begin{abstract}

%\end{abstract}
\begin{figure}[b]
\centering
\includegraphics[width=3.5cm]{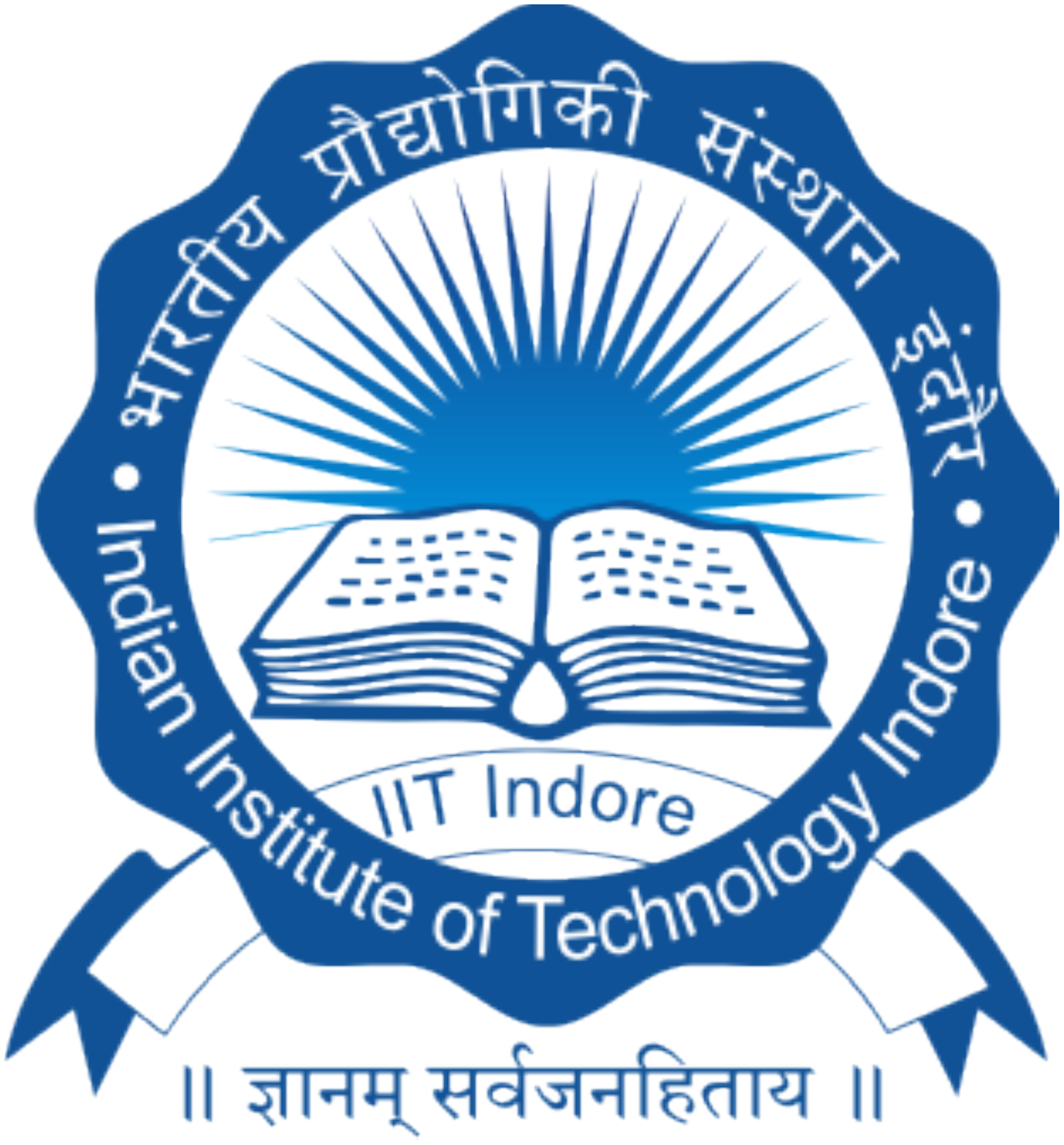}
\end{figure}

% insert suggested PACS numbers in braces on next line
%\pacs{25.75.Nq, 25.75.Dw}
% insert suggested keywords - APS authors don't need to do this
%\keywords{constituent quarks, multi-strange baryons, QGP, strangeness scaling}

%\maketitle must follow title, authors, abstract, \pacs, and \keywords
\maketitle 

% body of paper here - Use proper section commands
% References should be done using the \cite, \ref, and \label commands
%\section{}
% Put \label in argument of \section for cross-referencing
%\section{\label{}}
%\subsection{}

%\maketitle \\original

\newpage
%\section{INTRODUCTION} 
\section{Why High Energies?}
\begin{figure}[h]
\centering
\includegraphics[width=5.5cm]{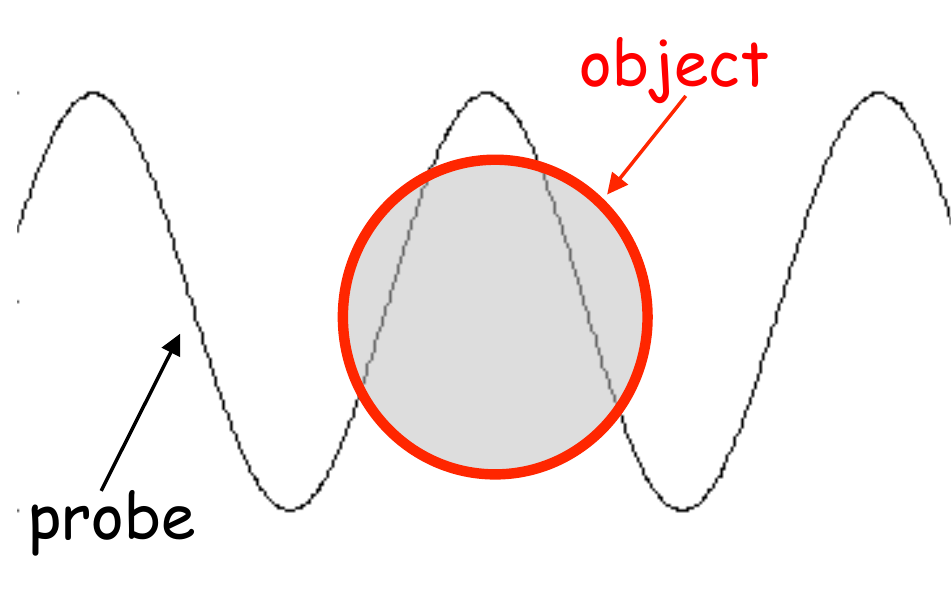}
\includegraphics[width=5.5cm]{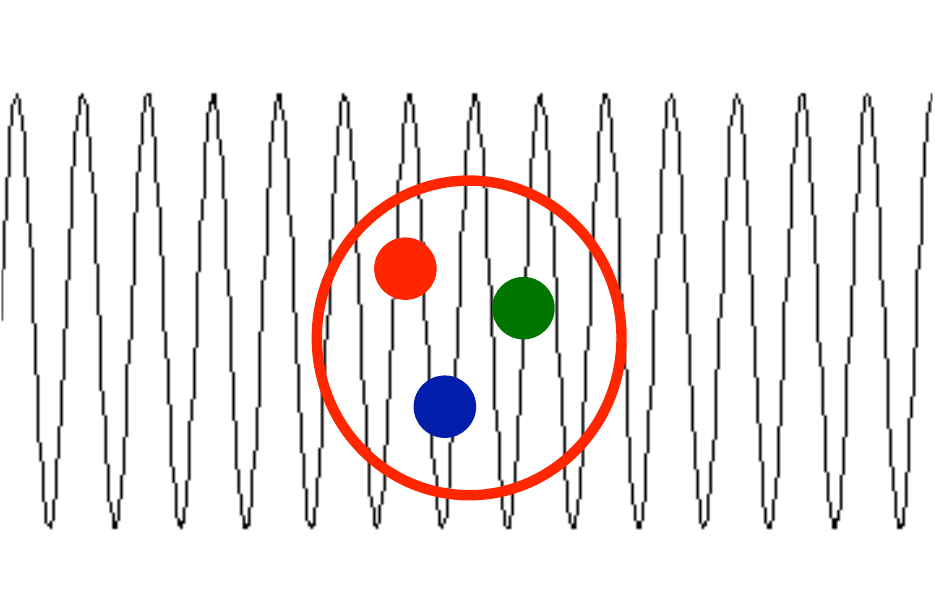}
\caption{Left: A low-energy probe probing an object, Right: a high-energy probe able to probe deep inside of the object because of higher resolution.}
\label{Fig:deBroglie}
\end{figure}

In Particle Physics, we deal with elementary constituents of matter. By elementary we mean the particles have no substructure or they are point-like objects. However, the elementariness depends on the spatial resolution of the probe used to investigate the possible structure/sub-structure \cite{perkins}. The resolution is $\Delta r$ if two points in an object can just be resolved as separate when they are a distance $\Delta r$ apart. Assuming that the probing beams themselves consist of point-like particles like electrons or positrons, the resolution is limited by the de Broglie wavelength of these beam particles, which is given by $\boxed{\lambda ~=~ h/p}$, where $p$ is the beam momentum and $h$ is Planck's constant. Hence beams of high momentum have short de Broglie wavelengths and can have high resolution. For example if we need to probe a dimension of  $1~Fermi$ ($10^{-15} ~ meter$) (let's say the inner structure of proton, the charge radius of proton being $\sim 0.840 ~ fm$), we need to use a beam of momentum $ 1.47~ GeV$ which is given by de Broglie's above expression.  Figure \ref{Fig:deBroglie} shows a schematic picture where a low energy probe fails to probe the inner structure of an object unlike a high energy probe with higher resolution. In addition to the above consideration, Einstein's formula $E~=~mc^2$ helps us to produce particles of higher masses (like the massive gauge bosons, Higgs etc.) in nature's way.

\section{Special Theory of Relativity \& Invariants}
\label{sec:1}
In particle physics, the particles are treated relativistically, meaning $E \approx pc \gg mc^2$ and thus special theory of relativity becomes an  mathematical tool in describing the particle kinematics.\\

\begin{itemize}
\item{ Space and time can not be treated independently as is done in Newtonian mechanics.}
\item{Physical objects that were treated as an independent three component vector and a scalar in non-relativistic physics mix in high-energy phenomena.}
\item{Combined to form a four-component Lorentz vector that transforms like a time and space coordinate.}  
\end{itemize}

For their consistent and unified treatment, one relies on Einstein's theory of special relativity (STR), having the following two underlying principles.\\

\begin{itemize}
\item{\underline {Invariance of velocity of light}:  velocity of light always remains as the constant $c$ in any {\it inertial frame}}.
\item{\underline{Relativity Principle:} This requires covariance of the equations, namely the physical law should keep its form invariant in any inertial frame of reference. In mathematical language this amounts to the fact that physical laws have to be expressed in Lorentz tensors.}
\end{itemize}
Note that the principle of relativity applies to Galilei transformation and is valid in Newtonian mechanics as well. But the invariance of the velocity of light necessitates Lorentz transformation in changing from one inertial system to another that are moving relative to each other with constant speed.

\subsection{Lorentz Transformation}
Consider a Lorentz boost in $x$-direction. Here, a particle at $(t, x, y, z,)$ in a coordinate frame $L$ is boosted to $(t^{\prime}, x^{\prime}, y^{\prime}, z^{\prime} )$ with velocity $v$. This statement is equivalent to changing to another coordinate frame $L^{\prime}$ which is moving in the $x$-direction at velocity $-v$. $L^{\prime}$  is assumed to coincide with $L$ at $t=t^{\prime}=0$. Then the two coordinates are related by the following equations:\\
\begin{eqnarray}
\begin{aligned}
t \rightarrow t^{\prime} =\frac{t+\left(v/c^2\right)x}{\sqrt{1-\left(v/c\right)^2}} \Rightarrow {x^{0}}^{\prime} &=& \gamma\left(x^0+\beta x\right), \\%\nonumber \\
x \rightarrow x^{\prime} =\frac{x+vt}{\sqrt{1-\left(v/c\right)^2}} \Rightarrow {x^{1}}^{\prime} &=& \gamma\left(\beta x^0+ x\right), \\%\nonumber \\
{x^{2}}^{\prime} &=& x^2, \\%\nonumber \\
{x^{3}}^{\prime} &=& x^3 
\label{lt}
\end{aligned}
\end{eqnarray}
where, $\beta = v/c$, $\gamma = \frac{1}{\sqrt{1-\beta^2}}$. The above equations can be written in matrix form as:\\
\begin{equation}
\left[ \begin{array}{c}  {x^{0}}^{\prime} \\ {x^{1}}^{\prime} \\ {x^{2}}^{\prime} \\{x^{3}}^{\prime} \end{array} \right]
=\begin{bmatrix} \gamma & \beta \gamma & 0 & 0 \\ \beta \gamma & \gamma & 0 & 0\\ 0 & 0 & 1 & 0 \\ 0 & 0 & 0 & 1 \\ \end{bmatrix}  \times \left[ \begin{array}{c} x^{0} \\ x^{1}\\ x^{2} \\x^{3} \end{array} \right] 
\end{equation}

\subsubsection{The Proper Time ($\tau$)}
{\bf It is the time an observer feels in the observer's rest frame.} \\

 Proper time  $\boxed{d\tau \equiv dt \sqrt{1-\beta^2}}$ is a Lorentz invariant scalar.\\
 \underline{Proof:}\\
 \begin{eqnarray}
 ds^2 &=& (cdt)^2 -dx^2 -dy^2 - dz^2 \nonumber \\
 &=& c^2dt^2\left[ 1- \frac{dx}{dt}^2-\frac{dy}{dt}^2-\frac{dz}{dt}^2 \right] \nonumber \\
 &=&c^2dt^2 \left(1-\beta^2 \right) \nonumber \\
 &=& \left(cd\tau \right)^2 \nonumber
 \end{eqnarray}
 is Lorentz invariant by definition.
\subsection{What is need of the variable called "Rapidity"?} 
Successive Lorentz boost in the same direction is represented by a single boost, where the transformation velocity is given by \\

$\beta^{\prime\prime} = {|{v/c}|}^{\prime \prime} = \frac{\beta+{\beta}^{\prime}}{1+\beta {\beta}^{\prime}}$\\

\underline{Proof:}\\
Assume velocity $v^{\prime}$ in frame $L$ is observed as $v^{\prime \prime}$ in frame $L^{\prime \prime}$, where the frame $L^{\prime}$ is travelling in the $x$-direction with $-v$ in frame $L$. The coordinates $(t^{\prime}, {x^1}^{\prime})$ are expressed in terms of $(t, x^1)$ using the usual Lorentz transformation equations given by Eqn. \ref{lt}. Omitting other coordinates for simplicity, one obtains:\\
\begin{eqnarray}
{x^{0}}^{\prime} &=& \gamma\left(x^0+\beta x^1\right) \nonumber \\
{x^{1}}^{\prime} &=& \gamma\left(\beta x^0+ x^1\right) \nonumber \\
\beta^{\prime} &=& \frac{v^{\prime}}{c} = \frac{dx^1}{dx^0} \nonumber
\end{eqnarray}
then 
\begin{eqnarray}
\beta^{\prime\prime} = \frac{d{x^1}^{\prime}}{d{x^0}^{\prime}} &=& \frac{\gamma\left(\beta dx^0+d x^1\right)} {\gamma\left(dx^0+\beta d x^1\right)} \nonumber \\
\Rightarrow \beta^{\prime\prime} &=& \frac{\beta + \beta^{\prime}}{1+\beta \beta^{\prime}}
\end{eqnarray}

\underline{The velocity is not an additive quantity.} {\it i.e.} {\bf non-linear in successive transformation.}
Here comes the need of {\it "Rapidity"} to circumvent this drawback, by defining \\

$\beta = tanh~y$ or $y = \frac{1}{2} ln \left(\frac{1+\beta}{1-\beta}\right)$.
One can show (we will be showing this in subsequent sections) that rapidity is an additive quantity {\it i.e.} \\

$\boxed{y^{\prime \prime} = y + y^{\prime}}$\\

Using the rapidity, a Lorentz transformation with finite $\eta$, can be decomposed into $N$ successive transformations with rapidity \\

$\Delta y = y/N$\\

Solving $\beta, ~ \gamma$ in terms of $y$, we have \\

$\beta ~=~ tanh ~y$, $\gamma ~=~ cosh ~y$, $\beta \gamma ~=~ sinh ~y$\\

Lorentz boost given by Eqn. \ref{lt} can be rewritten as \\
\begin{equation}
  \begin{aligned}
  {x^{0}}^{\prime} &= (cosh ~y)~ x^0 ~ +~ (sinh ~y)~ x^1,   \\
  {x^{1}}^{\prime} &= (sinh ~y) ~x^0 ~ +~ (cosh ~y)~ x^1,   
  \end{aligned}
  \label{etaRot}
\end{equation}
Comparing this with rotation in the $x-y$ plane:\\
\begin{equation}
  \begin{aligned}
  {x}^{\prime} &= xcos ~\theta ~ -~ ysin ~\theta,   \\
  {y}^{\prime} &= xsin ~\theta ~ +~ ycos ~\theta,   \\
  \end{aligned}
  \label{rot}
\end{equation}
Eqn. \ref{etaRot} can be obtained from Eqn. \ref{rot} by substituting\\
\begin{itemize}
\item {$\theta  \rightarrow  -iy$} (this $y$ is rapidity variable )
\item {$x  \rightarrow  ix^0$}
\item{ $y  \rightarrow  x^1$} (this $y$ is the Cartesian coordinate )
\end{itemize}
Lorentz boost (in the $x$-direction) is formally a rotation by an angle $(-iy)$ in the $x$ and imaginary time $(ix^0)$ plane.\\

{\bf Experimental Consideration:} In high-energy collider experiments, the secondary particles which are produced from the interaction, are boosted in the $z$-direction (along the beam axis). The boosted angular distribution is better expressed as rapidity distribution.  At high-energies, each particle has $E \sim pc$, $p_{II} = p cos~\theta$, and its rapidity is approximated by so-called pseudo-rapidity:
\begin{equation}
  {\eta}^{\prime} = \frac{1}{2}~ln \left(\frac{1+\beta_{II}}{1-\beta_{II}}\right)  ~=~ \frac{1}{2}~ln \left(\frac{E+p_{II}c}{E-p_{II}c}\right) ~ \sim -ln ~ tan~\theta /2.
\end{equation}
This fact is taken into account in designing detectors, which are divided into modules that span the same solid angle in the $\eta-\phi$ (azimuthal angle) plane.

%\subsection{ The Invariant Line element}

\subsection{Four Vectors}
\begin{figure}[h]
\centering
\includegraphics[width=8.5cm]{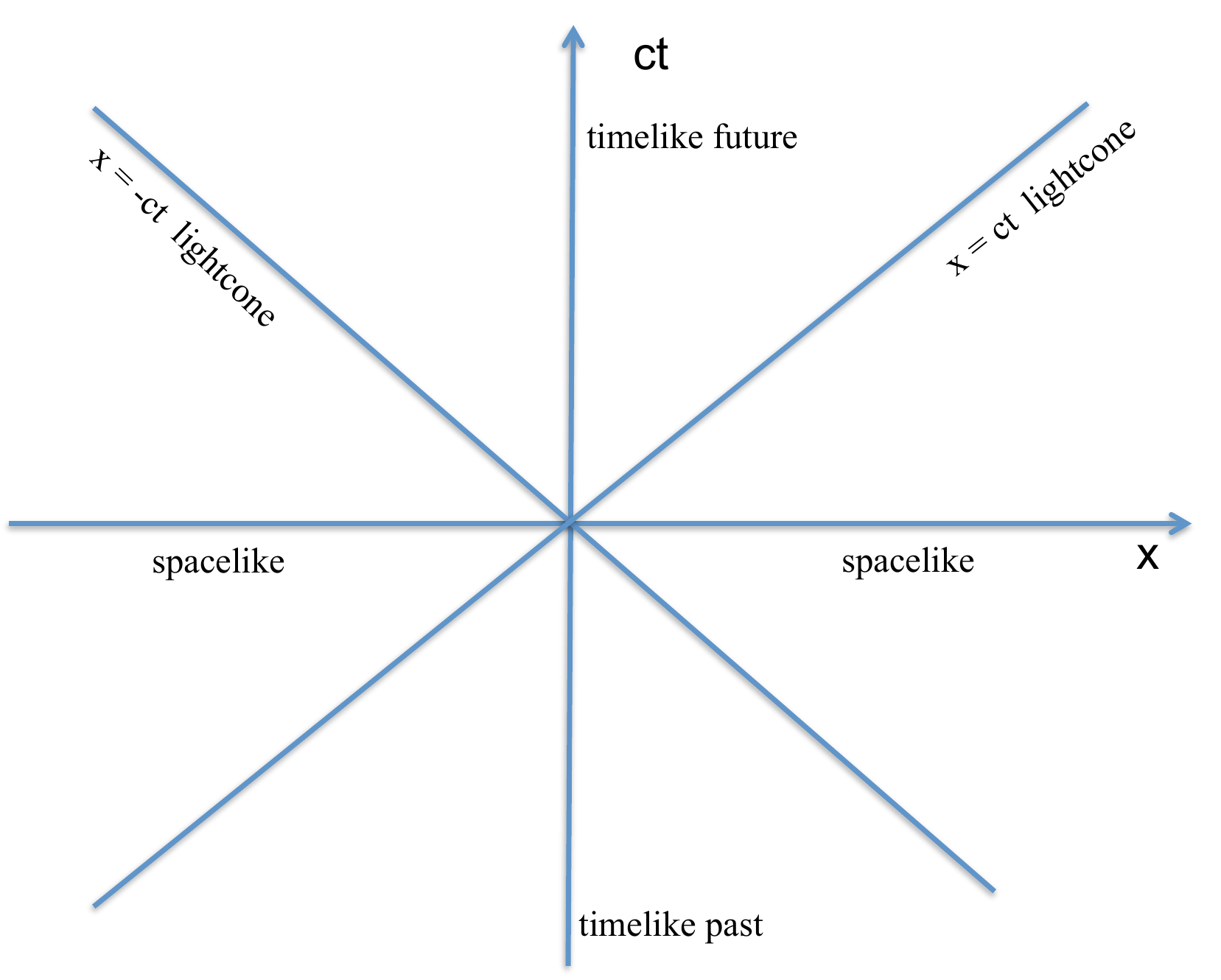}
\caption{A schematic of lightcone diagram.}
\label{Fig:lightcone}
\end{figure}

Th position-time 4-vector: $x^{\mu}$, $\mu ~=~ 0, 1,2,3$; with $x^0 ~=~ ct$, $x^1 ~=~ x$, $x^2 ~=~ y$, $x^3 ~=~ z$. 
\begin{eqnarray}
I ~ &\equiv& ~ (x^0)^2 - (x^1)^2 - (x^2)^2 -(x^3)^2  \nonumber \\
&=&  ({x^0}^{\prime})^2 - ({x^1}^{\prime})^2-({x^2}^{\prime})^2-({x^3}^{\prime})^2
\end{eqnarray}
$I$ is called the 4-dimensional length element, which is Lorentz Invariant (LI). A quantity having same value in all inertial frames is called an \underline{"invariant"}. This is like $r^2 ~=~ x^2 + y^2 + z^2$ being invariant under spatial rotation. $I$ could be written in the form of a sum:\\

\begin{equation}
I ~=~ \sum_{\mu =0}^3 ~ x^{\mu}x^{\mu} 
\end{equation}
To take care of the negative signs, let's define a "metric" $g_{\mu \nu}$ such that\\
\begin{equation}
g ~=~\begin{bmatrix} 1 & 0 & 0 & 0 \\ 0 & -1 & 0 & 0\\ 0 & 0 & -1 & 0 \\ 0 & 0 & 0 & -1 \\ \end{bmatrix}  
\end{equation}
Now
\begin{equation}
I ~=~ g_{\mu \nu} ~ x^{\mu}x^{\nu} 
\end{equation}
Define covariant 4-vector $x_{\mu}$ (index down):\\
\begin{equation}
x_{\mu} ~=~ g_{\mu \nu} ~ x^{\nu} 
\end{equation}
$x^{\mu}$ (index up) is called "contravariant 4-vector". With the above definitions, now
\begin{equation}
I ~=~ x_{\mu}x^{\mu} ~=~ x^{\mu}x_{\mu} 
\end{equation}
To each contravariant 4-vector $a^{\mu}$, a covariant 4-vector could be assigned and vice-versa.
\begin{eqnarray}
a^{\mu} ~&=&~ g^{\mu \nu} a_{\nu} \\
a_{\mu} ~&=&~ g_{\mu \nu} a^{\nu} 
\end{eqnarray}
$g^{\mu \nu}$ are the elements in $g^{-1}$. Since $g^{-1} ~=~ g$, 
$\boxed{g^{\mu \nu} ~=~ g_{\mu \nu}}$.\\
Given any two 4-vectors, $a^{\mu}$ and $b^{\mu}$,

\begin{equation}
a^{\mu}b_{\mu} ~=~  a_{\mu}b^{\mu} ~=~ a^0b^0 -a^1b^1 -a^2b^2 - a^3b^3
\end{equation}
is $L.I.$. The above operation is called "\underline{4-vector scalar product}". Remember Einstein's summation convention (repeated Greek indices are to be summed).

\begin{eqnarray}
a.b &\equiv&   a_{\mu}b^{\mu} \nonumber \\
&=& a^0b^0 -\vec{a}.\vec{b}
\end{eqnarray}

\begin{equation}
a^2 ~\equiv ~  a.a  ~=~ (a^0)^2 -{\vec{a}}^2
\end{equation}
The 1st term is called "{\it temporal}" and the 2nd is called {\it spatial component}.
\begin{itemize}
\item{If $a^2  ~>~ 0$: $a^{\mu}$ is called {\it time-like}. Events are in the forward light-cone. They appear later than the origin, $O$. Events in the backward light-cone appear earlier to $O$. Only events in the backward light-cone can influence $O$. And $O$ can have an influence only on the events in the forward cone.}
\item{If $a^2  ~<~ 0$: $a^{\mu}$ is called {\it space-like}. Events are called space-like events and there is no interaction with $O$. This is related to "{\it causality}".}
\item{If $a^2  ~=~ 0$: $a^{\mu}$ is called {\it light-like}. Connects all those events with the origin which can be reached by a light signal.}
\end{itemize}
This is shown pictorially in Figure \ref{Fig:lightcone}.
\subsection{Energy-Momentum Four-Vector}
The velocity of a particle is given by 
\begin{equation}
\vec{v} ~=~ \frac{d\vec{x}}{dt}
\end{equation}
where $d\vec{x}$ is the distance travelled in the laboratory frame and $dt$ is the time measured in the same frame. Proper velocity of the particle is given by
\begin{equation}
\vec{\eta} ~=~ \frac{d\vec{x}}{d\tau}
\end{equation}
where $d\vec{x}$ is the distance travelled in the laboratory frame and $d\tau$ is the proper time. Now
\begin{eqnarray}
\vec{\eta} ~=~ \frac{d\vec{x}}{d\tau} &=& \frac{d\vec{x}}{dt} \frac{dt}{d\tau} \nonumber \\
&=& \vec{v} \gamma \nonumber \\
&\Rightarrow & \boxed{\vec{\eta} = \gamma \vec{v}}
\end{eqnarray}
It is easy to work with the proper velocity, $\vec{\eta}$, as only $d\vec{x}$ transforms under Lorentz transformation. Furthermore,
\begin{equation}
\eta^{\mu} ~=~ \frac{dx^{\mu}} {d\tau}
\end{equation} 
so

\begin{equation}
\eta^0 ~=~ \frac{dx^0} {d\tau} ~=~ \frac{d(ct)}{\frac{1}{\gamma}dt} ~=~ \gamma c
\end{equation} 
Hence
\begin{equation}
\boxed{\eta^{\mu} ~=~ \gamma\left( c, ~v_x, ~v_y, ~v_z\right)}
\end{equation} 
This is called the {\it proper velocity 4-vector}. Remember that the spatial component brings up the negative sign for covariant tensor. Now
\begin{eqnarray}
\eta^{\mu} \eta_{\mu} &=& \gamma^2\left( c^2 -v_x^2-v_y^2-v_z^2 \right) \nonumber \\
&=& \gamma^2 c^2 \left( 1- \frac{v^2} {c^2} \right) \nonumber \\
&=& \gamma^2 c^2 \left( 1- \beta^2 \right) \nonumber \\
&=& c^2,
\end{eqnarray}
which is Lorentz Invariant. This also proves that 4-vector scalar product is L.I. \\

We know $momentum ~=~ mass ~\times ~ velocity$. And velocity can be "ordinary velocity" or "proper velocity". Classically, both are equal (non-relativistic limit).
\underline{If $\vec{p} ~=~ m\vec{v}$, the conservation of momentum is inconsistent with the principle of relativity.} In relativity, momentum is the product of mass and proper velocity.
\begin{equation}
\vec{p} \equiv m \vec{\eta}
\end{equation}

\begin{equation}
p^{\mu} ~=~ m \eta^{\mu}
\end{equation}
The spatial component of $p^{\mu}$ constitutes the (relativistic) momentum 4-vector:
\begin{equation}
\vec{p} ~=~\gamma m \vec{v} ~=~ \frac{m\vec{v}}{\sqrt{1-v^2/c^2}} 
\end{equation}
\begin{equation}
\boxed{p^0 ~=~ \gamma mc}
\end{equation}
Relativistic energy, $E$:
\begin{equation}
E ~\equiv ~ \gamma mc^2 ~=~ \frac{mc^2}{\sqrt{1-v^2/c^2}} 
\end{equation}
Hence,
\begin{equation}
p^0 ~=~ \frac{E}{c}
\end{equation}
and the energy-momentum 4-vector:
\begin{equation}
\boxed{p^{\mu} ~=~ \left( \frac{E}{c}, ~ p_x, ~p_y, ~p_z \right)}
\end{equation}
Now
\begin{eqnarray}
p^{\mu} p_{\mu} &=& \frac{E^2}{c^2} -{\vec{p}}^{~2} ~=~ m^2c^2 : L.I. \nonumber \\
&=& \left(m\eta^{\mu}\right)\left(m\eta_{\mu}\right) \nonumber \\
&=& m^2 \left(\eta^{\mu} \eta_{\mu}\right) \nonumber \\
&=& m^2c^2 \nonumber \\
&\Rightarrow &\boxed{ p^{\mu} p_{\mu} = m^2} ~ (Natural ~Units).
\end{eqnarray}
\begin{equation}
\Rightarrow E^2 ~=~ {\vec{p}}^{~2} c^2 + m^2 c^4  \nonumber
\end{equation}
In natural units,
\begin{equation}
\boxed{E^2 ~=~ {\vec{p}}^{~2} + m^2} 
\end{equation}

%************************************
\begin{itemize}
\item{$p^2 ~=~ m^2 ~>~ 0$: Ordinary massive particle}
\item{$p^2 ~=~ m^2 ~=~ 0$: Massless particles like photons, gravitons etc.}
\item{$p^2 ~<~ 0$: Tachyon or virtual particles}
\item{$p^{\mu} ~=~0 $: Vacuum}
\end{itemize}

Remember that the relativistic equations $\vec{p} ~=~ \gamma m \vec{v}$ and $E ~=~ \gamma m$ do not  hold good for massless particles and $m~=~0$ is allowed only if the particle travels with the speed of light.
For massless particles,\\

$v ~=~ c$ and $E ~=~ |\vec{p}| c$.

\subsection{The Choice of Units}
We know
\begin{eqnarray}
x^2 &=& c^2t^2  -x_1^2-x_2^2-x_3^2 \\
p^2 &=& m_0^2 c^2
\end{eqnarray}
The velocity of light {\it "c"} appears directly in these and many other formulas. Furthermore,  
de Broglie relation between 4-momentum and wave vector of a particle is
\begin{equation}
E ~=~ \hbar \omega   ~~ (obtained ~from ~Einstein's ~equation)
\end{equation}
In 4-vector notation,
\begin{equation}
P ~=~ \hbar K
\end{equation}
where $P ~=~ \left\lbrace \frac{E}{c}, ~ p \right\rbrace$, $K ~=~ \left\lbrace \frac{\omega}{c}, ~ k \right\rbrace$. If we choose a system of unit in which\\

$\boxed{\hbar ~=~c =~ 1}$,\\

where $\hbar ~=~ \frac{h}{2\pi} ~=~ 1.055 \times 10^{-34} ~ Joule ~sec$: unit of action/angular momentum ($ML^2/T$).\\

$c ~=~ 2.998 \times 10^8 ~ meter ~sec^{-1}$: unit of velocity, the velocity of light in vacuum ($L/T$).\\

Now the relativistic formula for energy, \\

\begin{equation}
E^2 ~=~ p^2c^2 ~+ m_0^2 c^4
\end{equation}
 in this new system of unit (called natural unit and more popularly used in high-energy (particle) physics) becomes
 
 \begin{equation}
\boxed{E^2 ~=~ p^2 ~+ m_0^2} 
\end{equation}

We can define system of units completely, if we specify the unit of energy ($ML^2/T^2$). In particle physics, unit of energy is $GeV$ ($1 ~ GeV ~=~ 10^9 ~ eV$). This choice is motivated by the rest mass of proton $\sim ~ 1 ~ GeV$. This gives rise to mass ($m$), momentum ($mc$), energy ($mc^2$) in $GeV$.
Length ($\frac{\hbar}{mc}$) and time ($\frac{\hbar}{mc^2}$) in $GeV^{-1}$.  \\

Taking the values of $\hbar ~=~ c ~=~ 1$, one obtains,\\

$\boxed{1 ~ sec ~=~ 1.52 \times 10^{24} ~GeV^{-1}}$ \\

$\boxed{1 ~ meter ~=~ 5.07 \times 10^{15} ~GeV^{-1}}$ \\

$1 ~ fermi ~\equiv ~ 1~ fm ~=~ 10^{-13}~ cm ~=~ 10^{-15} m$\\

$\Rightarrow  \boxed{~ 1 ~fm ~=~ 5.07 ~ GeV^{-1}}$\\

$\boxed{1~ fm ~=~ 3.33 \times 10^{-23}~ sec} $\\

$\boxed{197 ~ MeV ~=~ 1 ~ fm ^ {-1}}$\\

Note: $1 ~ TeV ~=~ 10^3 ~ GeV ~=~ 10^6 ~ MeV ~=~ 10^9 ~ KeV ~=~ 10^{12} ~ eV$.\\

The additional advantage of using natural unit in high energy particle physics is that we deal with strong interaction, whose life time $\sim ~ 10^{-23} ~ sec$, the decay length of particle can be better expressed in terms of fermi.

\subsection{Collider Vs Fixed Target Experiment}
\subsubsection{For Symmetric Collisions ($A+A$)}
Consider the collision of two particles. In LS, the projectile with momentum 
${\bf p}_1$, energy $E_1$ and mass $m_1$ collides with a particle of mass 
$m_2$ at rest. The 4-momenta of the particles are\\
$p_1 ~=~ (E_1,{\bf p}_1),~~~~~~~~~~$ ~$p_2 ~=~ (m_2, {\bf 0})$ \\
In CMS, the momenta of both the particles are equal and opposite, the 4-momenta 
are\\

$p_1^* ~=~ (E_1^*,{\bf p}_1^*),~~~~~~~~~~$ ~$p_2^* ~=~ (E_2^*, -{\bf p}_1^*)$ \\

The total 4-momentum of the system is a conserved quantity in the collision.\\

In CMS,
\begin{eqnarray}
p_{\mu}p^{\mu}=(p_1+p_2)^2 &=& (E_1+E_2)^2 - ({\bf p}_1+{\bf p}_2)^2 \nonumber \\ 
&=& (E_1+E_2)^2 \nonumber \\ 
&=& E_{cm}^2 ~\equiv ~s 
\end{eqnarray}
$\sqrt{s}$ is the total energy in the CMS which is the invariant mass of the CMS.\\

In LS, 
\begin{equation}
p_{\mu}p^{\mu}=(p_1+p_2)^2 ~=~ m_1^2+m_2^2 + 2E_1m_2
\end{equation}
Hence 
\begin{equation}
\boxed{E_{cm} ~=~ \sqrt{s} ~=~ \sqrt{m_1^2+m_2^2 + 2E_{proj}m_2}}
\end{equation}
where $E_1 = E_{proj}$, the projectile energy in LS. Hence it is evident here 
that the CM frame with an invariant mass $\sqrt{s}$ moves in the laboratory in 
the direction of ${\bf p}_1$ with a velocity corresponding to:\\
Lorentz factor, 
\begin{eqnarray}
\gamma_{cm} &=& \frac{E_1+m_2}{\sqrt{s}}  \\
\Rightarrow \sqrt{s} &=& \frac{E_{lab}}{\gamma_{cm}}, 
\end{eqnarray}
this is because $E~=~ \gamma m$
and 
\begin{equation}
y_{cm} ~=~ cosh^{-1} ~\gamma_{cm}.
\end{equation}
The center of mass or center of momentum frame (CM/CMS) is at rest and the total momentum is zero. This makes it a suitable choice for solving kinematics problems.\\

%\begin{note}
\underline{\bf{Note:}}
We know that for a collider with head-on collision ($\theta = 180^0$)
\begin{equation}
s~=~E_{cm}^2 ~=~ m_1^2 + m_2^2 + 2(E_1.E_2+|{\bf p}_1||{\bf p}_2|) 
\end{equation}
For relativistic collisions,  $m_1, ~m_2 ~\ll~ E_1,E_2$
\begin{equation}
E_{cm}^2 ~\simeq ~ 4E_1E_2
\end{equation}
For two beams crossing at an angle $\theta$, 
\begin{equation}
E_{cm}^2 ~= ~ 2E_1E_2 (1+cos~\theta)
\end{equation}
The CM energy available in a collider with equal energies ($E$) for new particle
production rises linearly with $E$ {\it i.e.} 
\begin{equation}
E_{cm} ~\simeq ~ 2E
\end{equation}
For a fixed-target experiment the CM energy rises as the square root of the 
incident energy:
\begin{equation}
E_{cm} ~\simeq ~ \sqrt{2m_2E_1}
\end{equation}
Hence the highest energy available for new particle production is 
achieved at collider experiments. For example, at SPS fixed-target experiment 
to achieve a CM energy of 17.3 AGeV the required incident beam energy is 
158 AGeV. \\

 {\bf Problem:} Suppose two identical particles, each with mass $m$ and kinetic energy $T$, collide 
head-on. What is their relative kinetic energy, $T^{\prime}$ ({\it i.e.} K.E. of one in the rest frame of the other). Apply this to an electron-positron collider, where K.E. of electron (positron) is 1 GeV. Find the 
K.E. of electron if positron is at rest (fixed target).  
Which experiment is preferred, a collider or a fixed-target experiment?
%\end{note}
%\begin{note}
\vspace{0.5cm}

\underline{\bf{Note:}}
Most of the times the energy of the collision is expressed in terms of 
nucleon-nucleon center of mass energy. In the nucleon-nucleon CM frame, two 
nuclei approach each other with the same boost factor $\gamma$. The 
nucleon-nucleon CM is denoted by $\sqrt{s_{NN}}$ and is related to the total 
CM energy by
\begin{equation}
\sqrt{s} ~= ~ A~\sqrt{s_{NN}}
\end{equation}
This is for a symmetric collision with number of nucleons in each nuclei as $A$.
The colliding nucleons approach each other with energy $\sqrt{s_{NN}}/2$ and 
with equal and opposite momenta. The rapidity of the nucleon-nucleon center 
of mass is\\
 $y_{NN} = 0$
and taking $m_1=m_2=m_p $, the projectile and target nucleons are at equal 
and opposite rapidities.
\begin{equation}
y_{proj}~=~ -y_{target} ~=~ cosh^{-1}~\frac{\sqrt{s_{NN}}}{2m_p} ~=~ y_{beam}.
\end{equation}
%\end{note}

%\begin{note}:
\underline{\bf{Note:}}
Lorentz Factor\\
\begin{figure}
\centering
\includegraphics[height=7.0cm]{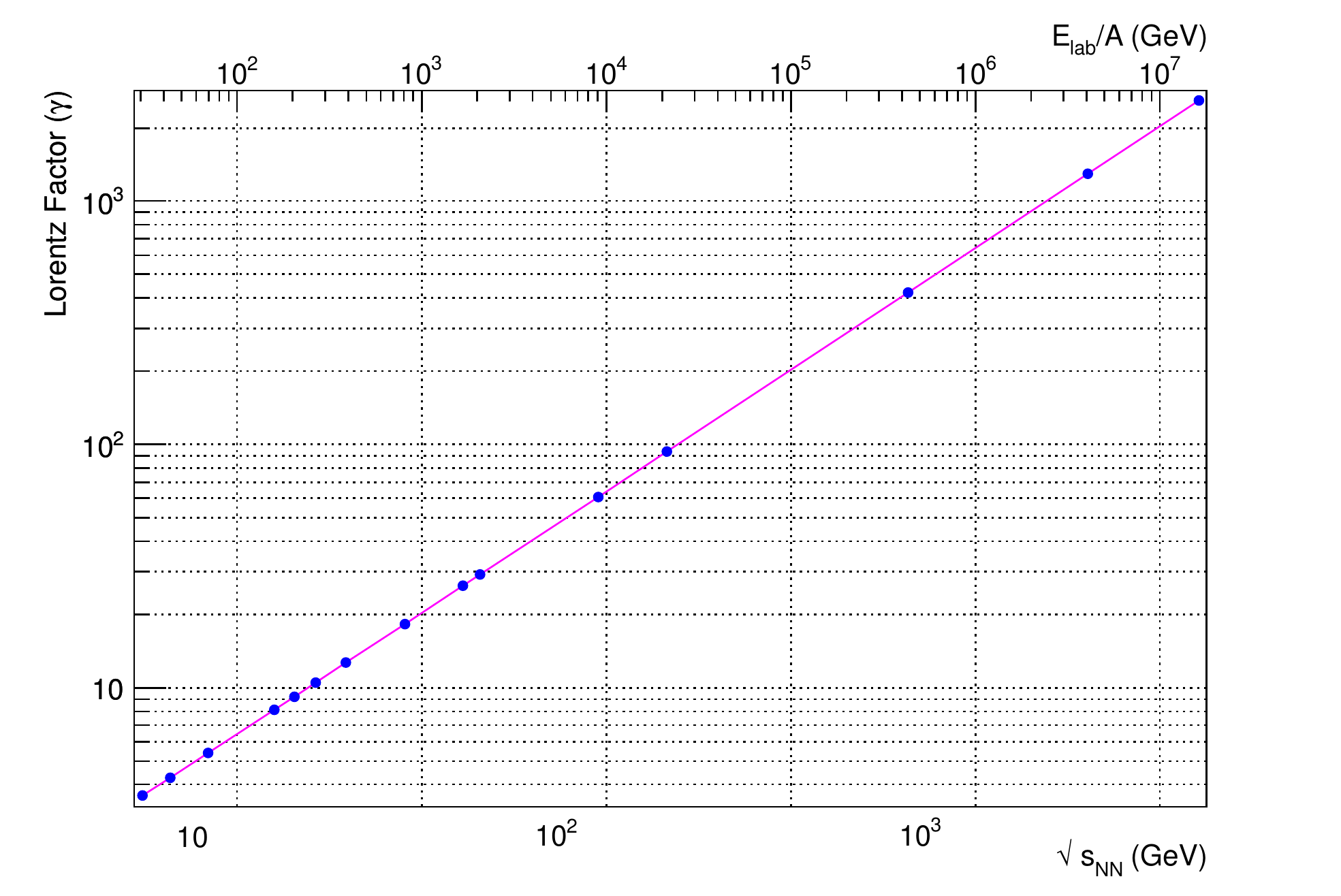}
\caption{A plot to demonstrate how the Lorentz Factor increases with
  collision energy for a symmetric collision. Center of mass system is
  compared with fixed
  target experiment. }
\label{LorentzFact}
\end{figure}

\begin{eqnarray}
\gamma &=& \frac{E}{M} \nonumber
=\frac{\sqrt{s}}{2A~m_p} \nonumber \\
&=&\frac{A~\sqrt{s_{NN}}}{2A~m_p} \nonumber
=\frac{\sqrt{s_{NN}}}{2~m_p} \\
&=&\frac{E_{beam}^{CMS}}{m_p} 
\end{eqnarray}
where E and M are Energy and Mass in CMS respectively. Assuming mass of a proton,
$m_p \sim 1$ GeV, the Lorentz factor is of the order of beam energy in CMS for a
symmetric collision.

%\end{note}

\subsubsection{For Asymmetric Collisions ($A+B$)}

During the early phase of relativistic nuclear collision research, the 
projectile mass was limited by accelerator-technical conditions ($^{38}$Ar 
at the Bevalac, $^{28}$Si at the AGS, $^{32}$S at the SPS). Nevertheless, 
collisions with mass $\approx$ 200 nuclear targets were investigated. 
Analysis of such collisions is faced with the problem of determining 
an "effective" center of mass frame, to be evaluated from the numbers 
of projectile and target participant nucleons, respectively. Their 
ratio - an thus the effective CM rapidity - depends on impact parameter. 
Moreover, this effective CM frame refers to soft hadron production only, 
whereas hard processes are still referred to the frame of nucleon-nucleon 
collisions. The light projectile on heavy target kinematics are described 
in \cite{asymmetric}.
 
The center of mass energy of a collision of two different systems with charge $Z_1$, $Z_2$ and atomic numbers $A_1$, $A_2$ with $Z~=~A~=~1$, for a proton is 
\begin{equation}
\boxed{\sqrt{s_{NN}} ~\simeq ~ 2 \sqrt{s_{pp}} ~+~ \sqrt{s_{pp}} \sqrt{\frac{Z_1Z_2}{A_1A_2}}} 
\end{equation}
where sub-index $NN$ refers to the energy per nucleon inside the colliding nucleus and
$\sqrt{s_{pp}}$ is the corresponding energy in $pp$ collisions. The rapidity shift in non-symmetric systems is given by 
\begin{equation}
\boxed{\Delta y ~\simeq ~ \frac{1}{2} ~ ln \left[\frac{Z_1A_2}{Z_2A_1}\right]}
\end{equation}
This is due to the fact that the center-of-mass frame of the $pA$ collision doesn't coincide with the laboratory center-of-mass frame. The rapidity shift in $p+Pb$ collision is
\begin{eqnarray}
\Delta y & \simeq  & \frac{1}{2} ~ ln \left[\frac{Z_1A_2}{Z_2A_1}\right] \nonumber \\
&=& \frac{1}{2} ~ ln \left[ \frac{82 \times 1}{1 \times 208}\right] \nonumber \\
&=& -0.465 \nonumber
\end{eqnarray}
Hence $\Delta y = \pm 0.465$ for $p+Pb$ collisions, flipping the beams. This rapidity shift need to be taken into account for the comparison with Pb+Pb data.

At LHC, the maximum proton beam energy is 7 TeV, while the maximum Pb beam energy is 2.75 TeV, $\sqrt{s} = 8.775$ TeV. The difference in available energy is due to the charge-to-mass ratio, $Z/A$. More is the number of neutrons in the nucleus, difficult it is to accelerate to higher energies. Because of different energies, the two beams will also not have the same rapidity. For the proton beam $y_p ~=~ 9.61$ and for the Pb beam it is $y_{Pb} ~=~ 8.67$. Thus the center of the collision is shifted away from $y_{cm} ~=~0$ by $\Delta y_{cm} ~=~ (y_p - y_{Pb})/2 = 0.47$.

\subsection{The Energy and Velocity of the Center of Momentum}
\begin{figure}
\centering
\includegraphics[height=7cm]{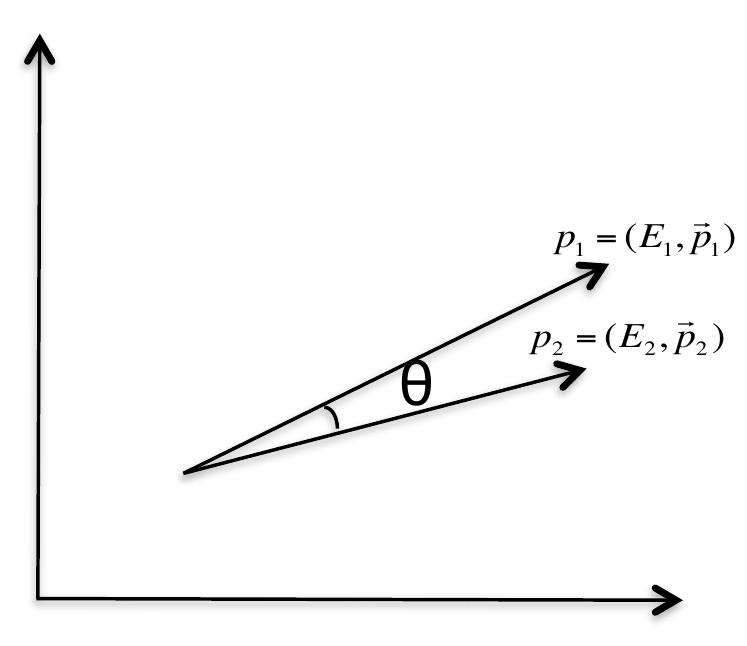}
\caption{The Laboratory system: two particle kinematics.}
\label{labFrame}
\end{figure}

As discussed in Ref.~\cite{hegedorn}, lets consider a Lorentz system- call it the laboratory system (frame), with two particles with masses, $m_1$ and $m_2$ and 
four momenta $p_1$ and $p_2$, respectively as shown in Fig. \ref{labFrame}.\\

What is the centre of mass energy, $E$ ?\\
\begin{enumerate}
\item It is independent of the Lorentz system, where  $p_1$ and $p_2$ are defined.

\item It must be possible to have an answer in terms of the three invariants, namely\\
$p_1^2 = m_1^2$ and $p_2^2 = m_2^2$ \\
and [$p_1p_2$ or $(p_1+p_2)^2$ or $(p_1-p_2)^2$].
\end{enumerate}
The answer would be trivial in the center of momentum (mass) frame itself. We use asterisk (*) in the CM frame to distinguish that from the laboratory frame. \\

In CM frame, 
\begin{eqnarray}
 (\vec{p_1}^*+\vec{p_2}^*) &=& 0 \nonumber \\
 \Longrightarrow p_1 + p_2 = (E_1^* + E_2^*, \vec{0}) \nonumber \\
 and ~E^* = E_1^* + E_2^*
 \end{eqnarray}
 
 Hence, \\
 \begin{eqnarray}
 (E^*)^2 &=& (E_1^* + E_2^*)^2 \nonumber \\
 &=& (p_1^* + p_2^*)^2  \nonumber \\
 &=& (p_1 + p_2)^2. 
 \end{eqnarray}

 This is because $(p_1 + p_2)^2$ is invariant.\\
 
 Let us define the total mass $M$ of the system as the square of the total 4-momentum as:\\
 
 $\boxed{M^2 = (p_1 + p_2)^2 = P^2 = (E^*)^2 \\
 = (E_1 + E_2)^2 - (p_1 + p_2)^2 = Invariant} $\\
 
Kinematically, for the given system, the two particles $p_1$ and $p_2$ are equivalent to one single particle with 4-momentum $P$ and mass $M=E_{\rm CM}$. Generalizing this, one can consider these individual particles to represent a system of particles.\\

Further, we know\\

$\boxed{\vec{p} = m\vec{v}\gamma, \rm{and}~~ \\
E = m\gamma}$, where  $\gamma = \frac{1}{\sqrt{1-\beta^2}}$.\\

Hence, the 4-momentum of the two-particle system is given by:\\

$\vec{P} = M\vec{\beta}\gamma$ \\

$E = M\gamma$.

Using the above, one obtains:\\
\begin{eqnarray}
\beta_{\rm CM} &=&\frac{\vec{P}}{E}  \\
&=& \frac{(\vec{p_1}+\vec{p_2})}{(E_1+E_2)}, 
\end{eqnarray}
is the velocity of the CM seen from the laboratory system.\\

\begin{eqnarray}
\gamma_{\rm CM} &=& \frac{1}{\sqrt{1-\beta^2}} \nonumber \\
&=& \frac{E}{M} \\
&=&\frac{E_1+E_2}{\sqrt{(E_1+E_2)^2-(\vec{p_1}+\vec{p_2})^2}}  \nonumber \\
&=&\frac{E_1+E_2}{E_{\rm CM}},
\end{eqnarray}

 is the Lorentz factor or the Lorentz boost of the CM. In general, {\it the Lorentz factor of the CM is the ratio of the sum of the energies of the particles in the laboratory system and the energy of the CM}.

\subsection{The Energy, Momentum and Velocity of one particle as seen from the rest system of another one}
Lets assume that we sit on particle $1$, which is in motion. What will be the energy of particle $2$, for us?\\

The answer to this question must always be the same, irrespective of the Lorentz system we start with. It is thus expressible by the invariants discussed in the preceding subsection (point-2), where the last two variables are the well-known Mandelstam's $s$ and  $t$ variables.\\

Let $E_{21}$: is the energy of particle 2, if we look at it sitting on particle $1$, which then appears to be at rest for us.\\
$E_{21} = E_2$, in the system where $\vec{p_1} =0$. One needs to write $E_{21}$ in terms of the invariants available in this problem.\\

Now, $p_1p_2 = E_1E_2-\vec{p_1}.\vec{p_2}$ = $m_1E_2$ (since $\vec{p_1} =0$).\\

Hence, $E_{21} = E_2 = \frac{p_1p_2}{m_1}$.\\

Note that the RHS is already in an invariant form.\\

$|\vec{p}_{21}|^2 = E_{21}^2 - m_2^2 = \frac{(p_1p_2)^2 - m_1^2m_2^2}{m_1^2}$.\\

If $p_1$ and $p_2$ are the momentum 4-vectors of any two particles in any Lorentz system, then\\

\begin{eqnarray}
E_{21} = \frac{p_1 p_2}{m_1} \nonumber \\
|\vec{p}_{21}|^2 = \frac{(p_1p_2)^2 - m_1^2m_2^2}{m_1^2}  ~~~~~(p_1p_2 \equiv E_1 E_2 -\vec{p}_1 \vec{p}_2) \nonumber \\
v_{21}^2 = \frac{|\vec{p}_{21}|^2}{E_{21}^2} = \frac{(p_1p_2)^2 - m_1^2m_2^2}{(p_1p_2)^2}
\label{eqn:Lab}
\end{eqnarray}

The above three equations give the energy, momentum and velocity of particle $2$, as seen from particle $1$. The velocity $v_{21}$ is the relative velocity, which is symmetric in $1$ and $2$. Note here that all these expressions are invariant and can be evaluated in any Lorentz system.

\subsection{The Energy, Momentum, and Velocity of a Particle as seen from the CM System}
This problem is now like all the above quantities are as seen from a {\it fictitious particle} $M$, called the ``{\it center-of-momentum-particle}", whose 4-momentum is\\
\begin{equation}
P= p_1 +p_2 
\label{total-P} 
\end{equation}

We need to apply formulae (Eq.~\ref{eqn:Lab}) with $p_1$ replaced by $P$ and $p_2$ by the 4-momentum of that particle whose energy, momentum and velocity, we like to determine. From Eq. \ref{eqn:Lab},
\begin{eqnarray}
E^*_{1} = \frac{P p_1}{M} \nonumber \\
|\vec{p}^*_{1}|^2 = \frac{(Pp_1)^2 - M^2 m_1^2}{M^2}  \nonumber \\
{v^*}_{1}^2 =  \frac{(Pp_1)^2 - M^2 m_1^2}{(Pp_1)^2}
\label{eqn:CM}
\end{eqnarray}
This is by using Eq.~\ref{total-P} and by using\\
\begin{eqnarray}
 p_1p_2 &=& \frac{1}{2}[(p_1+p_2)^2-p_1^2-p_2^2] \nonumber \\
 &=& \frac{1}{2} (M^2-m_1^2-m_2^2),
 \end{eqnarray}
 one obtains
\begin{eqnarray}
 E^*_1 = \frac{M^2 + (m_1^2-m_2^2)}{2M}  \\
 E^*_2 = \frac{M^2 - (m_1^2-m_2^2)}{2M}  \\
 E^*_1 + E^*_2 = M
 \end{eqnarray}

\begin{eqnarray}
 |\vec{p}^{~*}|^2 &=& |{\vec{p}^{~*}}_1|^2 = |{\vec{p}^{~*}}_2|^2 \nonumber  \\
  &=& \frac{M^4 - 2M^2(m_1^2+m_2^2) + (m_1^2-m_2^2)}{4M^2}  \nonumber \\
  &=& \frac{[M^2 -(m_1+m_2)^2 ]~[M^2- (m_1-m_2)^2]}{4M^2}  
 \end{eqnarray}

\begin{equation}
{v_1^{~*}}^2 = \left(\frac{|\vec{p}^{~*}|}{E^*_1}\right)^2.
\end{equation}
Here, $E^*_1, ~v_1^{~*}$ are energy and velocity of particle $1$, as seen from their common CM system and 
$M^2 =P^2 =(p_1+p_2)^2$ is the square of the total mass. The above equations give the energy, momenta and the velocities of two particles $m_1$ and $m_2$ for which

\begin{equation}
M \rightarrow m_1 + m_2 \nonumber
\end{equation}

\subsection{Description of Nucleus-Nucleus Collisions in terms of Light-Cone 
Variables}

In relativistic nucleus-nucleus collisions, it is convenient to use kinematic
variables which take simple forms under Lorentz transformations for the change
of frame of reference. A few of them are the light cone variables $x_+$ and 
$x_-$, the rapidity and pseudorapidity variables, $y$ and $\eta$. A particle 
is characterized by its 4-momentum, $p_{\mu}=(E,{\bf p})$. In fixed target 
and collider experiments where the beam(s) define reference frames, boosted 
along their direction, it is important to express the 4-momentum in terms of 
more practical kinematic variables.

\begin{figure}
\centering
\includegraphics[height=8cm]{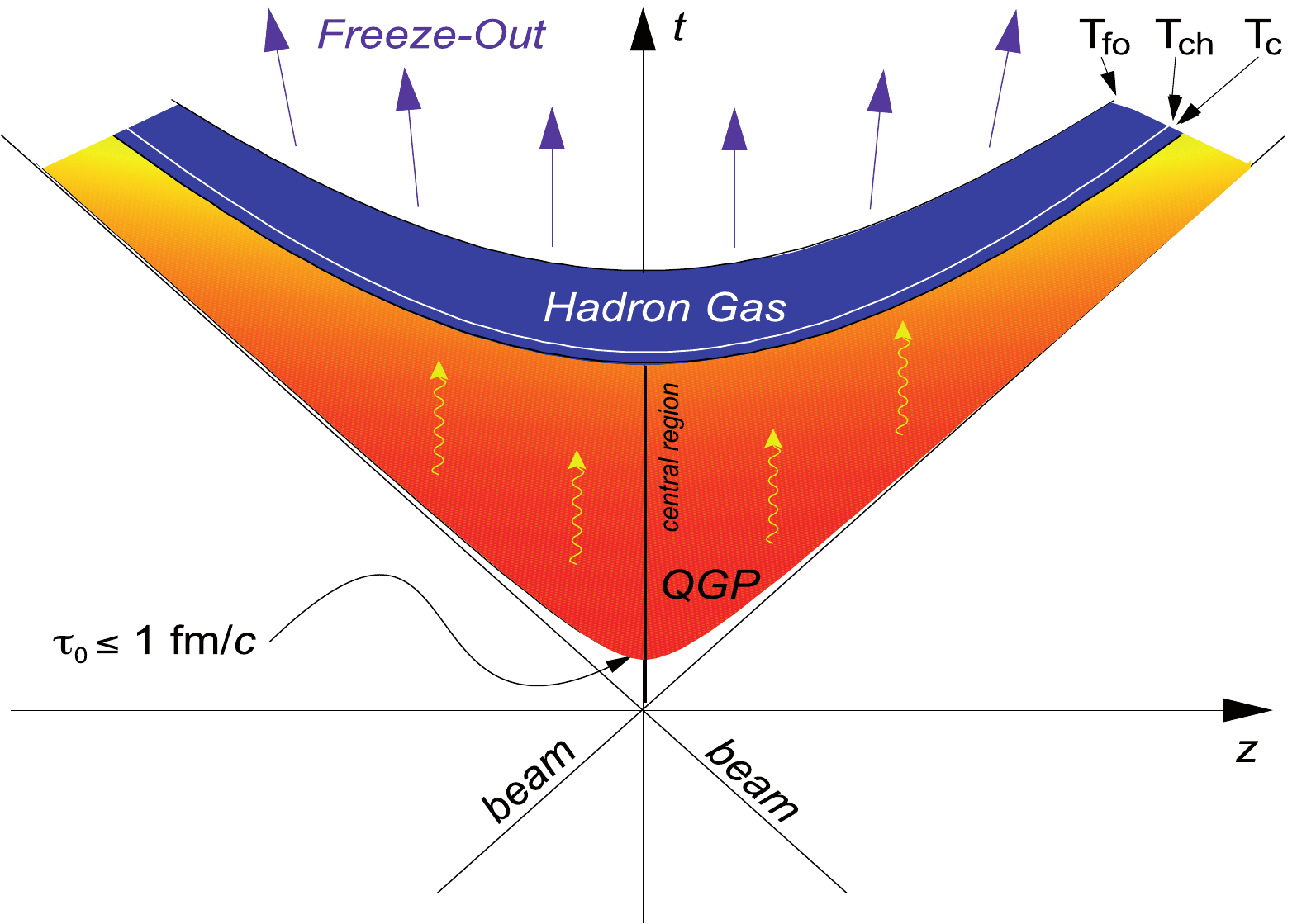}
% If not, use
%\picplace{5cm}{2cm} % Give the correct figure height and width in cm
%
\caption{Description of heavy-ion collisions in one space ($z$) and one time
($t$) dimension.}
\label{lightcone}
\end{figure}

%---------------------------------------------------------------------
Figure \ref{lightcone} shows the collision of two Lorentz contracted nuclei
approaching each other with velocities nearly equal to velocity of light.
The vertical axis represents the time direction with the lower half representing
time before the collision and the upper half, time after the collision. The
horizontal axis represents the spatial direction. Both the nuclei collide 
at $(t,z) = (0,0)$ and then the created fireball expands in time going through
various processes till the created particles freeze-out and reach the detectors.
The lines where $t^2 - z^2 ~=~ 0$ (note that $\sqrt{t^2 - z^2} ~\equiv~ \tau$, 
 $\tau$ being the proper time of the particle) along the path of the colliding 
nuclei define the light cone. The upper part of the light-cone, where 
$t^2 - z^2 ~ >~ 0$, is
the time-like region. In nucleus-nucleus collisions, particle production occurs
in the upper half of the $(t,z)$-plane within the light-cone. The region 
outside the light cone for which $t^2 - z^2 ~ <~ 0$ is called space-like region.
The space-time rapidity is defined as
\begin{equation}
\eta_s ~=~ \frac{1}{2}~ln\left(\frac{t+z}{t-z}\right)
\end{equation}
It could be seen that $\eta_s$ is not defined in the space-like region. It takes
the value of positive and negative infinity along the beam directions for which
$t = \pm z$ respectively. A particle is "light-like" along the beam direction. Inside the light-cone which
is time-like, $\eta_s$ is properly defined.

For a particle with 4-momentum $p~(p_0,{\bf p_T},p_z)$, the light-cone momenta
are defined by
\begin{eqnarray}
p_+ &=& p_0 + p_z\\
p_- &=& p_0 - p_z
\end{eqnarray}
$p_+$ is called ``{\it forward light-cone momentum}'' and $p_-$ is called
``{\it backward light-cone momentum}''.\\For a particle travelling along the
beam direction, has higher value of forward light-cone momentum and travelling
opposite to the beam direction has lower value of forward light-cone momentum.
The advantages of using light-cone variables to study particle production
are the following.\\
1. The forward light-cone momentum of any particle in one frame is related to
the forward light-cone momentum of the same particle in another boosted Lorentz
frame by a constant factor.\\
2. Hence, if a daughter particle $c$ is fragmenting from a parent particle
$b$, then the ratio of the forward light-cone momentum of $c$ relative to that
of $b$ is independent of the Lorentz frame.\\
Define
\begin{eqnarray}
x_+ &=& \frac{p_0^c + p_z^c}{p_0^b + p_z^b}\\
&=& \frac{c_+}{b_+} . \nonumber
\end{eqnarray}
The forward light-cone variable $x_+$ is always positive because $c_+$ can't be
greater than $b_+$. Hence the upper limit of $x_+$ is $1$. $x_+$ is Lorentz
invariant. \\
3. The Lorentz invariance of $x_+$ provides a tool to measure the momentum
of any particle in the scale of the momentum of any reference particle. 

\subsection{Pictorial Representation of Detector System}
\begin{figure}[h]
\centering
\includegraphics[width=10.5cm]{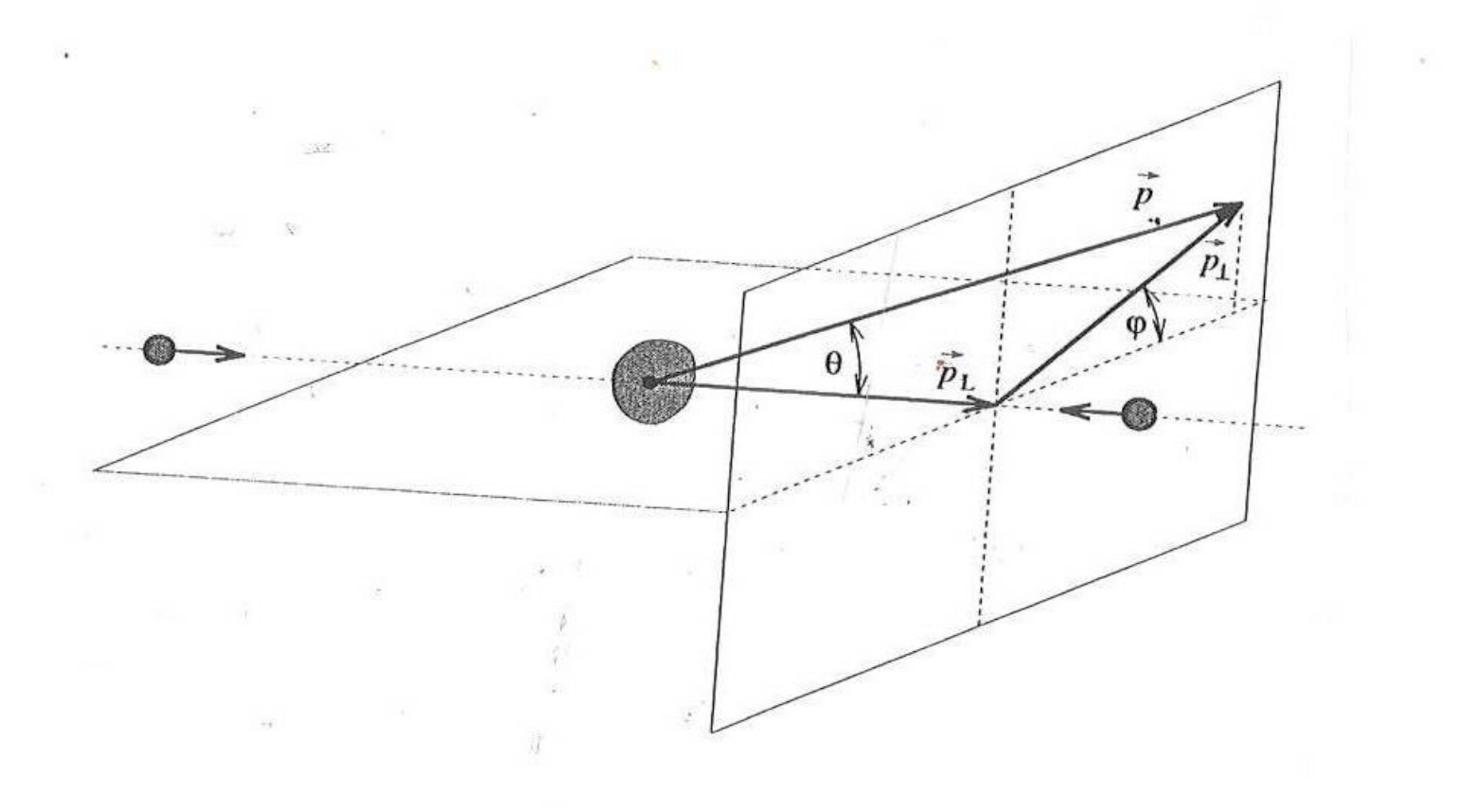}
\caption{A schematic decomposition of particle momentum $\vec{p}$ (in CM frame) into parallel and longitudinal components. Note the angle of inclination $\theta$ of $\vec{p}$ and the azimuthal angle $\phi$ of $p_{\perp}$ \cite{rafelskiBook}.}
\label{Fig:collisionPic}
\end{figure} 

In an collider experimental environment, a particle is emitted from
the collision point making a polar angle $\theta$ with the collision
point. When the momenta of the particles are determined by a tracking
detector, the (pseudo)-rapidity and azimuthal angles are given by\\
$y=tanh^{-1}
v_z=tanh^{-1}\frac{p_z}{E}=tanh^{-1}\frac{p_z}{\sqrt{p_x^2+p_y^2+p_z^2+m_0^2}}$\\

and $\boxed{\phi = tan^{-1}\frac{p_y}{p_x}}$.\\

The polar angle $\theta$ is given by\\
$\boxed{\theta ~=~ cos^{-1}\frac{p_z}{|\vec{p}|}~=~tan^{-1}\frac{|\vec{p_T}|}{p_z}}$. \\

Pictorially, these are shown in the Figure \ref{Fig:collisionPic}. A detector plane is spanned by
($\eta,\phi$), with $\eta$ decreasing while going away from the beam
axis in annular rings and $\phi$ is scanned making an angle with the
beam axis and increasing it anti-clock-wise, as is shown in the picture.

\subsection{The Rapidity Variable}
One of the important tasks in hand is to introduce key kinematic variables that relate particle momentum to the dynamics that is occurring in heavy-ion collisions. It is essentially the convenience of working in center-of-momentum system, we need to introduce the observable rapidity. To do so, let's proceed as follows:
The co-ordinates along the beam line (conventionally along the $z$-axis) is 
called {\it longitudinal} and perpendicular to it is called {\it transverse} 
({\it x-y}). The 3-momentum can be decomposed into the longitudinal ($p_z$) 
and the transverse (${\bf p}_T)$, ${\bf p}_T$ being a vector quantity which is 
invariant under a Lorentz boost along the longitudinal direction. The 
variable rapidity ``$y$'' is defined by
\begin{eqnarray}
y = \frac{1}{2}ln\left(\frac{E+p_z}{E-p_z}\right) \\
= ln \bigg(\frac{E+p_z}{m_T}\bigg)
\label{rapidityDef}
\end{eqnarray}
It is a dimensionless quantity related to the ratio of forward light-cone to 
backward light-cone momentum. The rapidity changes by an additive constant 
under longitudinal Lorentz boosts.
 
For a free particle which is on the mass shell (for which $E^2=p^2+m^2$), 
the 4-momentum has only three degrees of freedom and can be represented by 
$(y,{\bf p}_T)$. $(E,{\bf p}_T)$ could be expressed in terms of 
$(y,{\bf p}_T)$ as
\begin{eqnarray}
E &=& m_T ~ cosh ~y\\
p_z &=& m_T ~ sinh ~y
\end{eqnarray}
$m_T$ being the transverse mass which is defined as
\begin{equation}
m_T^2 = m^2+{\bf p}_T^2.
\end{equation}
The advantage of rapidity variable is that the shape of the rapidity 
distribution remains unchanged under a longitudinal Lorentz boost. When we go 
from CMS to LS, the rapidity distribution is the same, with the $y$-scale 
shifted by an amount equal to $y_{cm}$. This is shown below.

\subsubsection{Rapidity of Center of Mass in the Laboratory System}
The total energy in the CMS system is $E_{cm}=\sqrt{s}$. The energy and 
momentum of the CMS in the LS are $\gamma_{cm}\sqrt{s}$ and $\beta_{cm}
\gamma_{cm}\sqrt{s}$ respectively. The rapidity of the CMS in the LS is 
\begin{eqnarray}
y_{cm} &=& \frac{1}{2} ~ ln \left[\frac{\gamma_{cm}\sqrt{s} + \beta_{cm}
\gamma_{cm}\sqrt{s}}{\gamma_{cm}\sqrt{s} - \beta_{cm}\gamma_{cm}\sqrt{s}} 
\right]  \nonumber\\
&=& \frac{1}{2} ~ ln \left[\frac{1+\beta_{cm}}{1-\beta_{cm}}\right]
\end{eqnarray}
It is a constant for a particular Lorentz transformation.

\subsubsection{Relationship between Rapidity of a particle in LS and 
rapidity in CMS}
The rapidities of a particle in the LS and CMS of the collision are 
respectively,
$y = \frac{1}{2}~ln\left(\frac{E+p_z}{E-p_z}\right)$ 
and $y^* = \frac{1}{2}~ln\left(\frac{E^*+p_z^*}{E^*-p_z^*}\right)$. 
For a particle travelling in longitudinal direction, the Lorentz
transformation of its energy and momentum components give
\begin{equation}\label{engyMomLT}
\begin{bmatrix}
E^* \\
P_L^* 
\end{bmatrix}
=
\begin{bmatrix}
\gamma & -\gamma \beta \\
-\gamma \beta & \gamma
\end{bmatrix}
\cdot
\begin{bmatrix}
E \\
P_L 
\end{bmatrix},
P_T^* = P_T
\end{equation}
where $P_L$ and $P_T$ are the longitudinal and transverse components
of $\vec{P}$, which are parallel and perpendicular to $\beta$, respectively.
Hence, the inverse Lorentz transformations on $E$ and $p_z$ give
\begin{eqnarray}
y &=& \frac{1}{2} ~ ln \left[\frac{\gamma(E^*+\beta p_z^*) + 
\gamma(\beta E^*+p_z^*)}
{\gamma(E^*+\beta p_z^*) - \gamma(\beta E^*+p_z^*)}\right]  \nonumber\\
&=& \frac{1}{2} ~ ln \left[\frac{E^*+p_z^*}{E^*-p_z^*}\right] 
+ \frac{1}{2} ~ ln \left[\frac{1+\beta}{1-\beta}\right]\\
\Rightarrow y &=& y^* + y_{cm}.
\end{eqnarray}
Hence the rapidity of a particle in the laboratory system is equal to the sum 
of the rapidity of the particle in the center of mass system and the rapidity 
of the center of mass in the laboratory system. It can also be state that the 
rapidity of a particle in a moving (boosted) frame is equal to the rapidity 
in its own rest frame minus the rapidity of the moving frame. In the 
non-relativistic limit, this is like the subtraction of velocity of the
moving frame. However, this is not surprising because, non-relativistically, 
the rapidity $y$ is equal to longitudinal velocity $\beta$. Rapidity is a 
relativistic measure of the velocity. This simple property of 
the rapidity variable under Lorentz transformation makes it a suitable choice to 
describe the dynamics of relativistic particles. The simple shape invariance nature of rapidity spectra brings its importance in the analysis of particle production in nuclear collisions. For instance, in fixed-target experiments, we can study particle spectra using $y$ as a variable without making an explicit transformation to the CM frame of reference and from the rapidity spectra, we deduce the point of symmetry corresponding to the CM rapidity. In symmetric collisions with fixed targets, the CM frame is located in the middle between the rapidities of the projectile and target {\it i.e.} $y_{cm} ~=~ y_{proj}/2$. In this case, the particle rapidity spectrum must be symmetric around $y_{cm}$. This allows for complementing measured particle spectra: if these are available for, {\it e.g.} $y \geq y_{cm}$, a reflection at the symmetry point $y_{cm}$ gives us the part of the spectrum with $y \leq y_{cm}$, for which an experimental measurement is absent.

\subsubsection{Relationship between Rapidity and Velocity}
Consider a particle travelling in $z$-direction with a longitudinal velocity 
$\beta$. The energy $E$ and the longitudinal momentum $p_z$ of the particle are 
\begin{eqnarray}
E &=& \gamma m\\
p_z &=& \gamma \beta m\\
p_z &=& \beta E
\end{eqnarray}
where $m$ is the rest mass of the particle. Hence the rapidity of the particle 
travelling in $z$-direction with velocity $\beta$ is
\begin{eqnarray}
y_{\beta} &=& \frac{1}{2} ~ ln \left[\frac{E+p_z}{E-p_z}\right] 
= \frac{1}{2} ~ ln \left[\frac{\gamma m + \gamma \beta m}{\gamma m - \gamma 
\beta m}\right] \nonumber\\
&=& \frac{1}{2} ~ ln \left[\frac{1+\beta}{1-\beta}\right] 
\end{eqnarray} 
Note here that $y_{\beta}$ is independent of particle mass. A particle
is said to be relativistic in nature if $\gamma >> 1,~ \beta \approx
1$ and $E \approx p$. In other words, the energy of the particle is
much higher than its rest mass i.e. $E >> m_0$ and hence the energy
and total momentum of the particle are comparable. In the non-relativistic 
limit when $\beta$ is small, expanding $y_{\beta}$ in terms of $\beta$ leads to
\begin{equation}
y_{\beta} = \beta + {\it O}(\beta^3)
\end{equation}
Thus the rapidity of the particle is the relativistic realization of its 
velocity.

\subsubsection{Beam Rapidity}
\begin{figure}
\centering
\includegraphics[height=7.0cm]{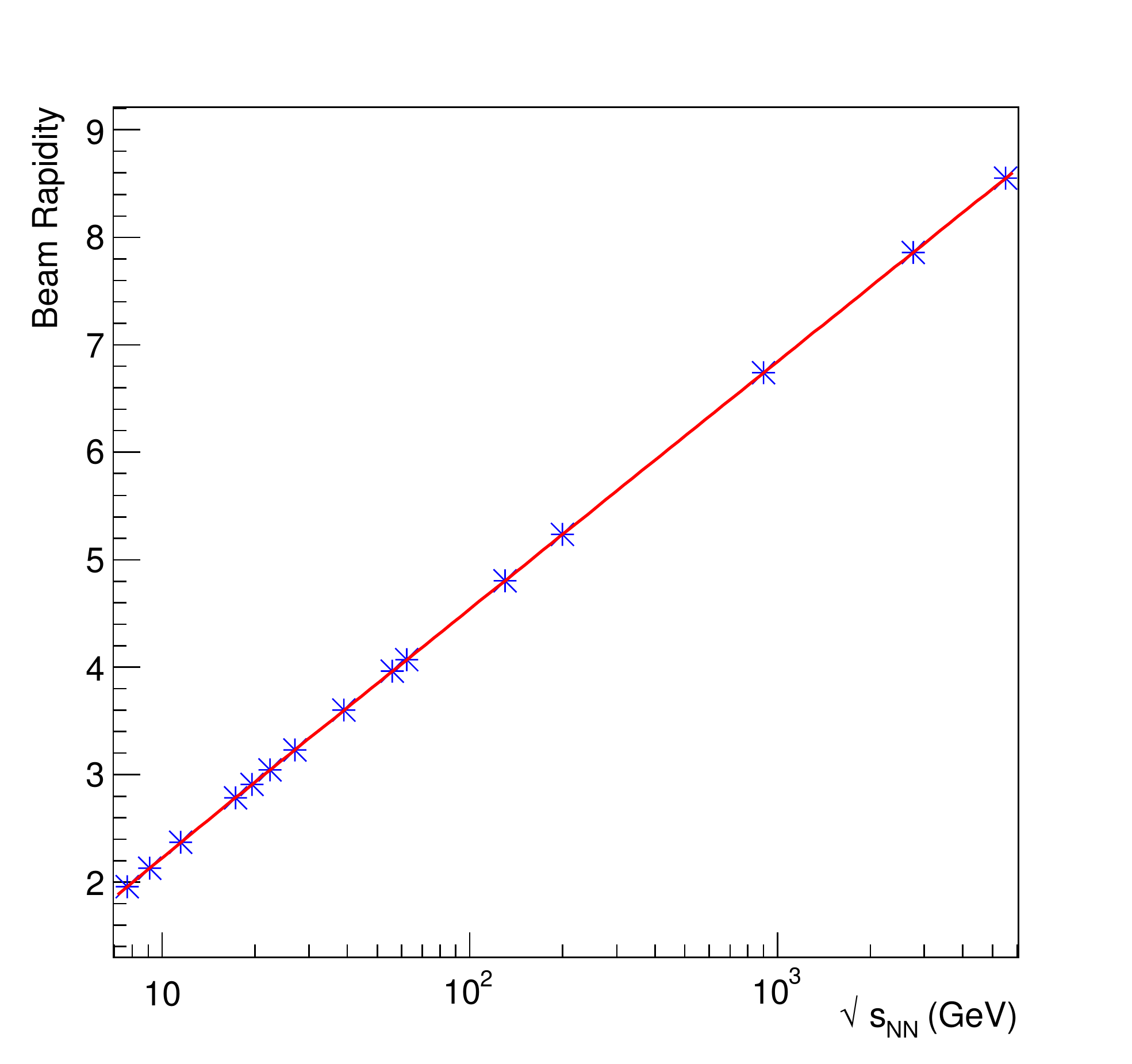}
\caption{A plot of beam rapidity as a function of center of mass energy.}
\label{beamRapE}
\end{figure}

We know, \\
$E ~=~ m_T ~ cosh ~y$, $p_z ~=~m_T ~ sinh ~y$ and $m_T^2 = m^2+{\bf p}_T^2 $.\\
For the beam particles, $p_T = 0$. \\
Hence, $E ~=~ m_b ~ cosh ~y_b$ and $p_z ~=~ m_b ~ sinh ~y_b$,\\
where $m_b$ and $y_b$ are the rest mass and rapidity of the beam particles.
\begin{equation}
\boxed{y_b = cosh^{-1} ~(E/m_b)} 
\label{eqn-yb-FT}
\end{equation}
Note that Eq. \ref{eqn-yb-FT} is used for fixed target experiment with
$E$ as the energy of individual beam particles, neglecting the rest mass. For example, for SPS
beam energy of 158 AGeV, $E=158$ GeV. This leads to $y_b = 5.82$ in fixed
target framework. Now when we convert Eq. \ref{eqn-yb-FT} for a
collider system, we get

\begin{eqnarray}
y_b &=& cosh^{-1} ~(E/m_b) \nonumber \\
&=& cosh^{-1} ~\left[\frac{\sqrt{s_{NN}}}{2~ m_p}\right] \\
&\Rightarrow & \boxed{y_b =~ ln(\sqrt{s_{NN}}/m_p)}
\label{eqn-yb-CM}
\end{eqnarray}

Now, for the above SPS beam energy of 158 AGeV, the equivalent center
of mass energy per nucleon is, $\sqrt{s_{\rm{NN}}} =$ 17.3 GeV. By
using Eq. \ref{eqn-yb-CM}, we get $y_b = 2.91$. This makes sense to
us, as in a collider experiment both the beams come in opposite
directions with equal beam rapidities.

Further,
\begin{equation}
 y_b = sinh^{-1} ~(p_z/m_b)
\end{equation}
In terms of the beam velocity, beam rapidity can be expressed as
\begin{equation}
 y_b = tanh^{-1} ~(p_z/E) = tanh^{-1} ~\beta
\end{equation}
Here $m_p$ is the mass of a proton. Note that the beam energy 
$E =\sqrt{s_{NN}}/2$ for a symmetric collider.\\ 

Furthermore, it could be shown that:
\boxed{y_b =~ \mp ln(\sqrt{s_{NN}}/m_p) ~=~ \mp y_{\rm{max}}}\\

%{\bf {\it Example:1}}\\
\underline{{\bf Example}}
For the nucleon-nucleon center of mass energy $\sqrt{s_{NN}} = 9.1$ GeV, 
the beam rapidity
$y_b ~=~ cosh^{-1} \left(\frac{9.1}{2 \times 0.938}\right)~=~2.26$\\
For p+p collisions with lab momentum 100 GeV/c,\\ 
$y_b ~=~ sinh^{-1} \left(\frac{p_z}{m_b}\right)
~=~sinh^{-1} \left(\frac{100}{0.938}\right)~=~5.36$\\
and for Pb+Pb collisions at SPS with lab energy 158 AGeV, $y_b ~=~ 2.91$.

% Needs checking for the PbPb collisions
 
\begin{table}[ht]
\caption{Table of collision energy, $\sqrt{s_{NN}}$ Vs the beam rapidity, $y_b$.}
\centering % used for centering table
\begin{tabular}{|c|c|c|c|c|c|c|c|c|c|c|c|c|c|c|c|c|} % centered columns (14 columns) separated by vertical bars
%\begin{longtable*}{c|c|c|c|c|c|c|c|c|c|c|c|c|c} % centered columns (14 columns) separated by vertical bars
%\hline\hline %inserts double horizontal lines
\hline % inserts single horizontal line
\mbox{$\sqrt{s_{NN}}$} (GeV) &7.7&9.1 &11.5& 17.3 & 19.6 &27.0&39.0        & 62.4    & 130   & 200  &900 & 2360 &2760   & 5520 & 7000 & 14000 \\ \hline 
\mbox{$y_b$}  &2.10 & 2.27 &2.50& 2.91 & 3.03&3.35&3.72  & 4.19 & 4.93 & 5.36 & 6.86 & 7.83 &7.98 & 8.68 & 8.91 & 9.61\\
\hline %inserts single line 
\label{table:beamRap}
\end{tabular}
\end{table}

\subsubsection{Rapidity of the CMS in terms of Projectile and Target Rapidities}
Let us consider the beam particle ``$b$'' and the target particle 
``$a$''.\\ $b_z ~=~ m_T~ sinh~y_b ~=~ m_b~sinh~y_b$. This is because $p_T$ 
of beam particles is zero. Hence
\begin{equation}
y_{b} ~=~ sinh^{-1}~ (b_z/m_b).
\end{equation}
The energy of the beam particle in the laboratory frame is \\
$b_0 ~=~m_T~cosh~y_b ~=~ m_b~cosh~y_b$.\\ 
Assuming target particle $a$ has longitudinal momentum $a_z$, its rapidity in the
laboratory frame is given by
\begin{equation}
y_{a} ~=~ sinh^{-1}~ (a_z/m_a)
\end{equation}
and its energy
\begin{equation}
a_{0} ~=~ m_a~cosh~y_a .
\end{equation}
The CMS is obtained by boosting the LS by  a velocity of the center-of-mass 
frame $\beta_{cm}$ such that the longitudinal momentum of the beam particle 
$b_z^*$ and of the target particle $a_z^*$ are equal and opposite. Hence 
$\beta_{cm}$ satisfies the condition,\\
$a_z^* ~=~ \gamma_{cm}(a_z-\beta_{cm}a_0) ~=~ -b_z^* ~=~ 
-\gamma_{cm}(b_z-\beta_{cm}b_0)$,
where $\gamma_{cm} ~=~ \frac{1}{\sqrt{1-\beta_{cm}^2}}$.
Hence,
\begin{equation}
\beta_{cm} ~=~ \frac{a_z+b_z}{a_0+b_0} .
\label{bcm}
\end{equation}
We know the rapidity of the center of mass is 
\begin{equation}
y_{cm} ~=~ \frac{1}{2}~ln \left[\frac{1+\beta_{cm}}{1-\beta_{cm}}\right]
\label{ycm}
\end{equation}
Using equations \ref{bcm} and \ref{ycm}, we get
\begin{equation}
y_{cm} ~=~ \frac{1}{2}~ln \left[\frac{a_0+a_z+b_0+b_z}{a_0-a_z+b_0-b_z}\right].
\end{equation}
Writing energies and momenta in terms of rapidity variables in the LS,
\begin{eqnarray}
y_{cm} &=& \frac{1}{2}~ln \left[\frac{m_a~cosh~y_a + m_a~sinh~y_a + 
    m_b~cosh~y_b + m_b~sinh~y_b}
  {m_a~cosh~y_a - m_a~sinh~y_a + m_b~cosh~y_b - m_b~sinh~y_b}\right] 
\nonumber \\
&=& \frac{1}{2} (y_a + y_b) ~+~ \frac{1}{2}~ln \left[ \frac{m_a~e^{y_a} + 
    m_b~e^{y_b}}
  {m_a~e^{y_b} + m_b~e^{y_a}}\right]
\end{eqnarray}
For a symmetric collision (for $m_a~=~m_b$),
\begin{equation}
  y_{cm} ~=~ \frac{1}{2}(y_a+y_b)
\end{equation}
Rapidities of $a$ and $b$ in the CMS are
\begin{equation}
y_a^* ~=~ y_a - y_{cm} ~=~ -\frac{1}{2} (y_b-y_a)
\end{equation}
\begin{equation}
y_b^* ~=~ y_b - y_{cm} ~=~ \frac{1}{2} (y_b-y_a).
\end{equation}
Given the incident energy, the rapidity of projectile particles and the 
rapidity of the target particles can thus be determined. The greater is the 
incident energy, the greater is the separation between the projectile and 
target rapidity.
\subparagraph{Central Rapidity} 
The region of rapidity mid-way between the projectile and target 
rapidities is called central rapidity.\\
\underline{{\bf Example}}
In p+p collisions at a laboratory momentum of 100 $GeV/c$, beam rapidity 
$y_b=5.36$, target rapidity $y_a=0$ and the central rapidity $\approx 2.7$.

\subsubsection{Mid-rapidity in Fixed target and Collider Experiments}
In fixed-target experiments (LS), $y_{target} ~=~ 0$.\\
$y_{lab} ~=~ y_{target} ~+ ~y_{projectile} ~=~y_{beam}$
Hence mid-rapidity in fixed-target experiment is given by,
\begin{equation}
y_{mid}^{LS} ~=~ y_{beam}/2 .
\end{equation}
In collider experiments (center of mass system),\\
$y_{projectile} ~=~ -y_{target} ~= ~y_{CMS} ~=~y_{beam}/2$.\\
Hence, mid-rapidity in CMS system is given by
\begin{equation}
y_{mid}^{CMS} ~=~ (y_{projectile}+y_{target})/2 ~=~ 0 .
\end{equation}
This is valid for a symmetric energy collider.
The rapidity difference is given by $y_{projectile}-y_{target}=2y_{CMS}$ and 
this increases with energy for a collider as $y$ increases with energy.

\subsubsection{Light-cone variables and Rapidity}
Consider a particle having rapidity $y$ and the beam rapidity is $y_b$. The
particle has forward light-cone variable $x_+$ with respect to the beam 
particle
\begin{eqnarray}
x_+ &=& \frac{p_0^c + p_z^c}{p_0^b + p_z^b}\nonumber \\
&=& \frac{m_T^c}{m^b}e^{y-y_b} 
\end{eqnarray}
where $m_T^c$ is the transverse mass of $c$. Note that the transverse momentum
of the beam particle is zero. Hence,
\begin{eqnarray}
y &=& y_b + ln~x_+ + ln\left(\frac{m_b}{m_T^c}\right)
\end{eqnarray}
Similarly, relative to the target particle $a$ with a target rapidity $y_a$, the
backward light-cone variable of the detected particle $c$ is $x_-$. $x_-$ is
related to $y$ by
\begin{equation}
x_- = \frac{m_T^c}{m^b}e^{y_a-y} 
\end{equation}
and conversely,
\begin{equation}
y = y_a - ln~x_- - ln\left(\frac{m_a}{m_T^c}\right).
\end{equation}
In general, the rapidity of a particle is related to its light-cone momenta by
\begin{equation}
y = \frac{1}{2}~ln\left(\frac{p_+}{p_-}\right)
\end{equation}
Note that in situations where there is a frequent need to work with boosts
along z-direction, it's better to use $(y,{\bf p_T})$ for a particle rather
than using it's 3-momentum, because of the simple transformation rules for
$y$ and ${\bf p_T}$ under Lorentz boosts.

\subsubsection{The Maximum Accessible Rapidity in an Interaction}
\begin{figure}[h]
\centering
\includegraphics[width=8.5cm]{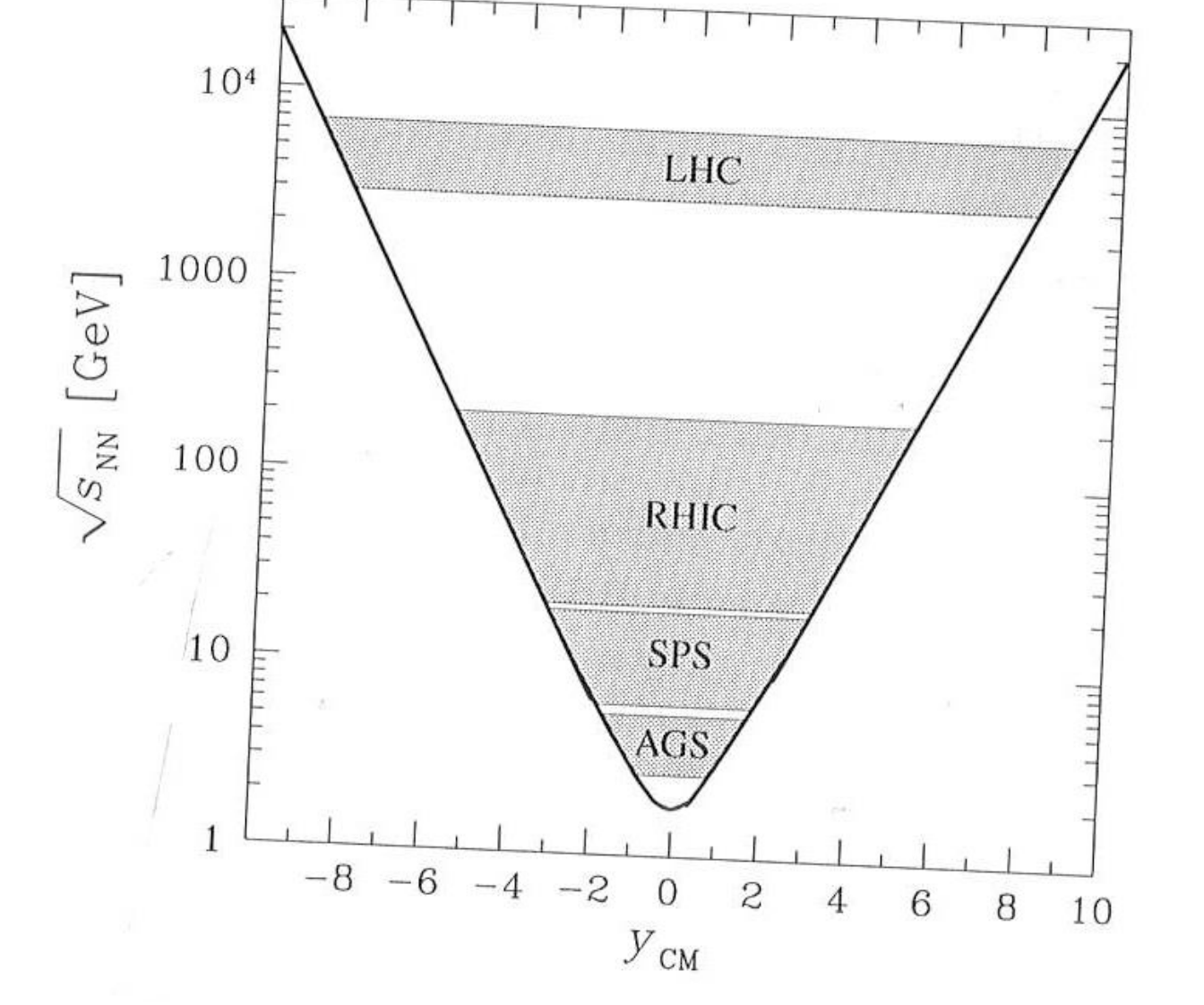}
\caption{$\sqrt{s_{NN}}$ (vertical axis) of various accelerators as a function of the projectile and target rapidities seen from the CM frame. Shaded areas: energy ranges accessible at various accelerators \cite{rafelskiBook}.}
\label{Fig:rapidityRange}
\end{figure}

We know the variable rapidity ``$y$'' of a particle is defined by Eqn. ~\ref{rapidityDef}:
\begin{equation}
y = \frac{1}{2}ln\left(\frac{E+p_z}{E-p_z}\right).
\end{equation}
This corresponds to 
\begin{equation}
tanh (y) = \frac{p_z}{E}
\end{equation}
 where, $p_z$ is the longitudinal momentum along the direction of the incident particle, 
$E$ is the energy, both defined for a given particle. The accessible range of rapidities 
for a given interaction is determined by the available centre-of-mass energy and all 
participating particles' rest masses. One usually gives the limit for the incident 
particle, elastically scattered at zero angle: 
\begin{eqnarray}
|y|_{max} &=& ln [(E+p_z)/m]  \nonumber \\
&=& ln [\gamma + \gamma \beta]  \nonumber \\
&=& ln [\gamma + \sqrt{\gamma^2-1}] \nonumber \\
&\simeq& cosh^{-1} \gamma, ~~ if ~ \gamma \gg 1
\end{eqnarray}
with all variables referring to the through-going particle given in the desired frame 
of reference (e.g. in the centre of mass). A Lorentz boost  $\beta$ along the direction 
of the incident particle adds a constant, $ ln [\gamma + \gamma \beta]$, to the rapidity. 
Rapidity differences, therefore, are invariant to a Lorentz boost. Statistical particle 
distributions are flat in $y$ for many physics production models. Note here that
$\frac{\partial y}{\partial p_z} = 1/E$. \\

Furthermore,
\begin{equation}
\gamma ~=~ \frac{E_{beam}}{m_p}
\end{equation}
and for symmetric collision,
\begin{equation}
E_{beam} ~=~ \frac{\sqrt{s_{NN}}}{2}
\Rightarrow ~ \gamma ~=~ \frac{\sqrt{s_{NN}}}{2m_p}
\end{equation}
Hence,
\begin{eqnarray}
y_{max} &=& cosh^{-1} ~ \left[\frac{\sqrt{s_{NN}}}{2m_p}\right] \nonumber \\
&=& cosh^{-1} ~ \left[\frac{E_{lab}}{Am_p}\right] \nonumber \\
& \Rightarrow & \boxed{y_{max}= y_b ~=~ ln \left[\frac{\sqrt{s_{NN}}}{m_p}\right]} 
\end{eqnarray}
Note that the maximum accessible rapidity is independent of the collision species for a symmetric collider and only depends on the center of mass energy.\\

{\bf Examples:}\\
(a) For RHIC top energy, $\sqrt{s_{NN}} ~= 200$ GeV, $\gamma = E_{beam}/m_p = 106.609$. Hence
$y|_{max} = 5.36$. For LHC, center of mass energy of 5.5 TeV ($\gamma = 2931.768$), $y|_{max} = 8.67$.\\

(b) $y_{beam}^{lab} ~=~ 5.8 $ for $E_{lab} ~=~ 158$ AGeV and 4.4 for $E_{lab} ~=~$ 40 AGeV.

\subsection{The Pseudorapidity Variable}
\begin{figure}
\centering
\includegraphics[height=7.0cm]{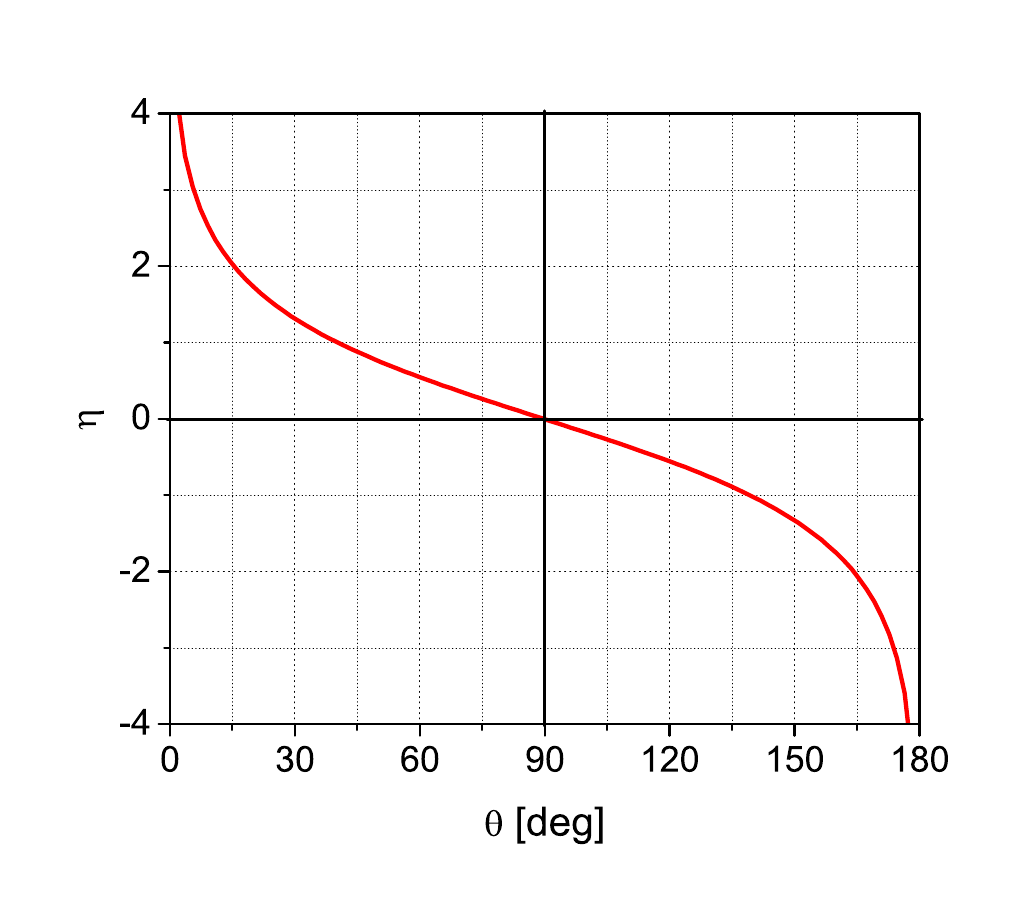}
\caption{A plot of pseudorapidity variable, $\eta$ as a function of the
polar angle, $\theta$. }
\label{etaDist}
\end{figure}

Let us assume that a particle is emitted at an angle $\theta$ relative to the 
beam axis. Then its rapidity can be written as\\
$y ~=~ \frac{1}{2}~ln \left(\frac{E+P_L}{E-P_L}\right) 
~=~ \frac{1}{2}~ln \left[\frac{\sqrt{m^2+p^2} + p~cos~\theta}{\sqrt{m^2+p^2} 
- p~cos~\theta}\right]$. 
At very high energy, $p \gg m$ and hence
\begin{eqnarray}
y &=& \frac{1}{2}~ln \left[\frac{p+p~cos~\theta}{p-p~cos~\theta}\right] 
\nonumber \\
  &=& -ln ~ tan~\theta/2 \equiv \eta
\end{eqnarray}
$\eta$ is called the pseudorapidity. 
Hence at very high energy,
\begin{equation}
y ~\approx ~ \eta ~=~ -ln ~ tan~\theta/2.
\end{equation}
In terms of the momentum, $\eta$ can be re-written as 
\begin{equation}
\eta ~=~ \frac{1}{2}~ln \left[\frac{|{\bf p}| + p_z}{|{\bf p}| - p_z}\right].
\label{eta}
\end{equation}
$\theta$ is the only quantity to be measured for the determination of 
pseudorapidity, independent of any particle identification mechanism. 
Pseudorapidity is defined for any value of mass, momentum and energy of the 
collision. This also could be measured with or without momentum information 
which needs a magnetic field. A plot of pseudorapidity as a function of the
polar angle, $\theta$ is shown in Fig. \ref{etaDist}. Table \ref{table:eta} shows
the values of $\eta$ corresponding to the polar angle of emission a particle.
One speaks of the "forward" direction in a  collider experiment, which refers 
to regions of the detector that are close to the beam axis, at high $|\eta |$.
The difference in the rapidity of two particles is independent of Lorentz boosts along the 
beam axis. Pseudorapidity is odd about $\theta = 90$ degrees. In other words, $\eta(\theta) = -\eta (180 - \theta)$.
\begin{table}[ht]
\caption{Table of pseudorapidity, $\eta$ Vs the polar angle, $\theta$.} % title of Table
\centering % used for centering table
\begin{tabular}{|c|c|c|c|c|c|c|c|c|c|c|c|c|c|c|} % centered columns (14 columns) separated by vertical bars
%\begin{longtable*}{c|c|c|c|c|c|c|c|c|c|c|c|c|c} % centered columns (14 columns) separated by vertical bars
%\hline\hline %inserts double horizontal lines
\hline % inserts single horizontal line
\mbox{$\theta$} & 0         & 5    & 10   & 20   & 30   & 45   & 60   & 80    & 90 & 100      & 110     & ... & 175     & 180 \\ \hline 
\mbox{$\eta$}   & $\infty$  & 3.13 & 2.44 & 1.74 & 1.32 & 0.88 & 0.55 & 0.175 & 0  & $-0.175$ & $-0.55$ & ... & $-3.13$ & -$\infty$ \\
\hline %inserts single line 
\end{tabular}
%\end{longtable*}
\label{table:eta} % is used to refer this table in the text
\end{table}

\begin{figure}
\centering
\includegraphics[height=6.0cm]{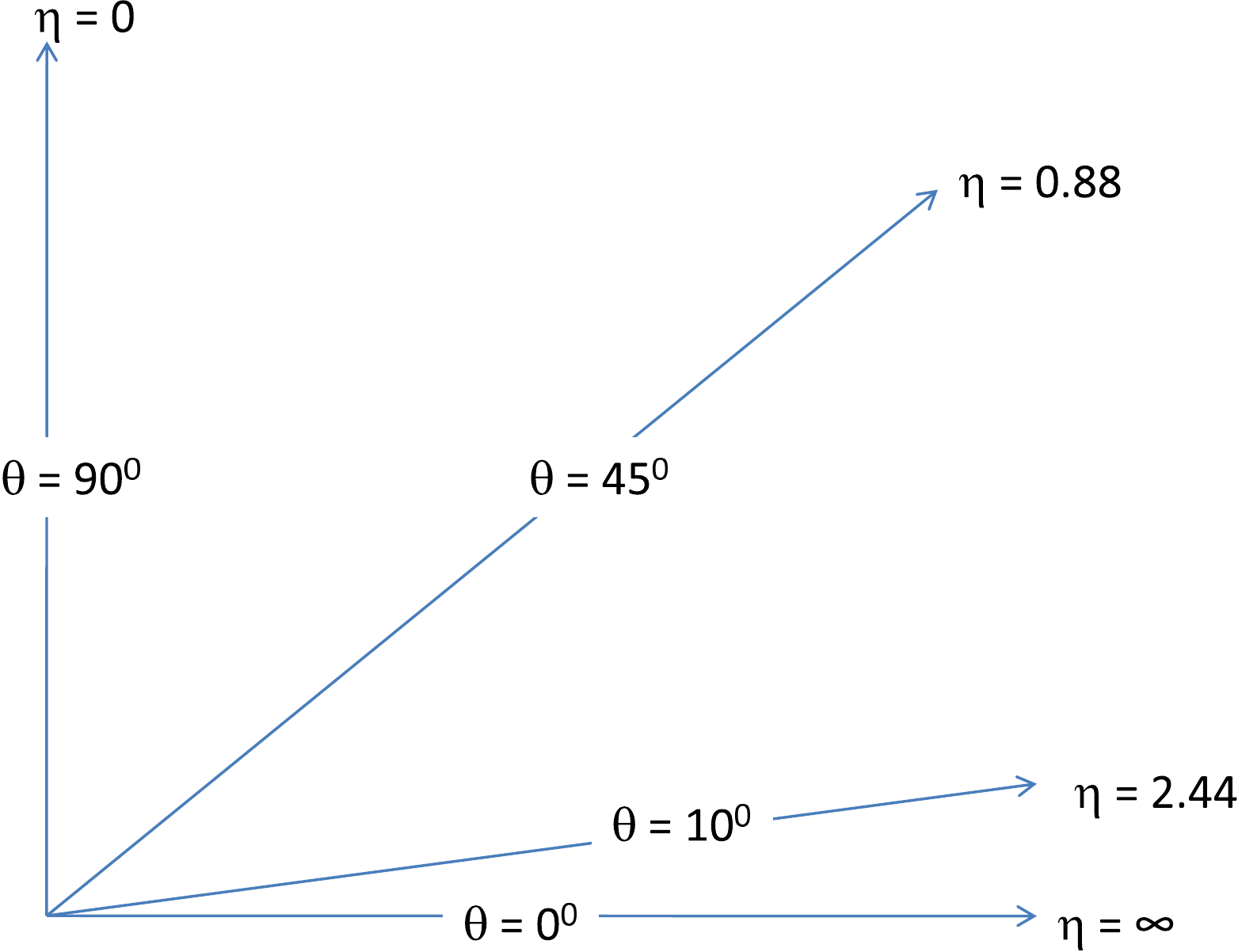}
\caption{ As angle increases from zero, pseudorapidity 
decreases from infinity.}
\label{etaSchem}
\end{figure}

\begin{figure}
\centering
\includegraphics[height=6cm]{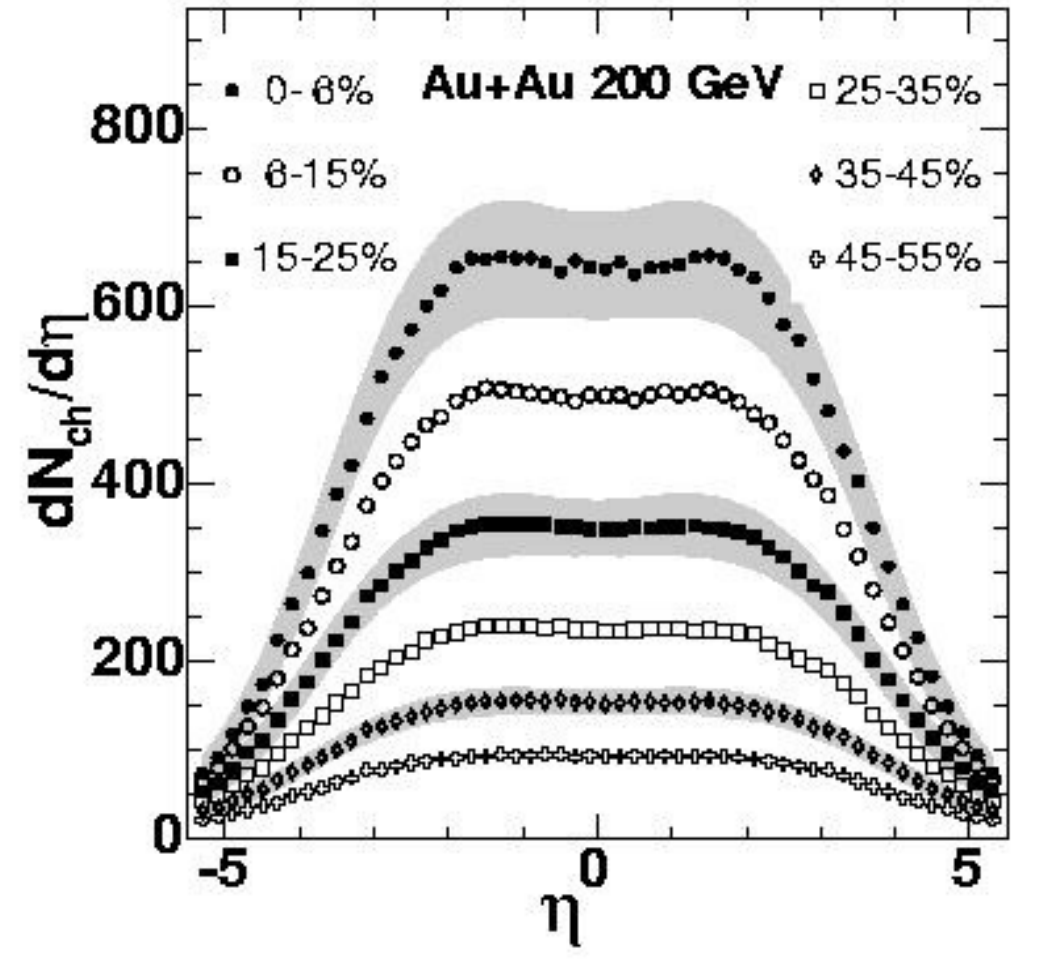}
\caption{The mid-rapidity $dN_{ch}/d\eta$ for Au+Au collisions at $\sqrt{s_{NN}}=$
200 GeV \cite{phobos}. }
\label{dndeta}
\end{figure}

\subsubsection{Change of variables from $(y,{\bf p}_T)$ to $(\eta,{\bf p}_T)$}
By equation \ref{eta},
\begin{eqnarray}
e^{\eta} &=& \sqrt{\frac{|{\bf p}| + p_z}{|{\bf p}| - p_z}}\\
e^{-\eta} &=& \sqrt{\frac{|{\bf p}| - p_z}{|{\bf p}| + p_z}}
\end{eqnarray}
Adding both of the equations, we get
\begin{equation}
|{\bf p}| ~=~p_T ~ cosh~\eta
\end{equation}
${\bf p}_T ~=~ \sqrt{|{\bf p}|^2-p_z^2}$. 
By subtracting the above equations, we get
\begin{equation}
p_z ~=~ p_T ~ sinh~\eta
\end{equation}
Using these equations in the definition of rapidity, we get
\begin{equation}
y ~=~ \frac{1}{2}~ln \left[\frac{\sqrt{p_T^2~cosh^2~\eta + m^2} + p_T~sinh~\eta}
{\sqrt{p_T^2~cosh^2~\eta + m^2} - p_T~sinh~\eta}\right]
\end{equation}
Similarly $\eta$ could be expressed in terms of $y$ as,
\begin{equation}
\eta ~=~ \frac{1}{2}~ln \left[\frac{\sqrt{m_T^2~cosh^2~y - m^2} + m_T~sinh~y}
{\sqrt{m_T^2~cosh^2~y - m^2} - m_T~sinh~y}\right]
\end{equation}
The distribution of particles as a function of rapidity is related to the 
distribution as a function of pseudorapidity by the formula
\begin{equation}
\frac{dN}{d\eta d{\bf p}_T} ~=~ \sqrt{1-\frac{m^2}{m_T^2~cosh^2~y}}~
\frac{dN}{dyd{\bf p}_T} .
\label{eqn:eta-y-conversion}
\end{equation}
In the region $y \gg 0$, the pseudorapidity distribution ($dN/d\eta$) and the 
rapidity distribution ($dN/dy$) which are essentially the ${\bf p}_T$-integrated
 values of $\frac{dN}{d\eta d{\bf p}_T}$ and $\frac{dN}{dy d{\bf p}_T}$ 
respectively, are approximately the same. 
In the  region $y \approx 0$, there is a small ``{\it depression}'' in 
$dN/d\eta$ distribution compared to $dN/dy$ distribution due to the above 
transformation. At very high energies where $dN/dy$ has a mid-rapidity plateau, 
this transformation gives a small dip in $dN/d\eta$ around $\eta \approx 0$
(see Figure \ref{dndeta}). However, for a massless particle like photon, the 
dip in $dN/d\eta$ is not expected (which is clear from the above equation). 
Independent of the frame of reference where $\eta$ is measured, the difference 
in the maximum magnitude of $dN/d\eta$ appears due to the above transformation. 
In the CMS, the maximum of the distribution is located at
$y \approx \eta \approx 0$ and the $\eta$-distribution is suppressed by a factor
$\sqrt{1-m^2/<m_T^2>}$ with reference to the rapidity distribution. In the 
laboratory frame, however the maximum is located around half of the beam 
rapidity $\eta \approx y_b/2$ and the suppression factor is 
$\sqrt{1-m^2/<m_T^2>~cosh^2~(y_b/2)}$, which is about
unity. Given the fact that the shape of the rapidity distribution is 
independent of frame of reference, the peak value of the pseudorapidity 
distribution in the CMS frame is lower than its value in LS. This suppression 
factor at SPS energies is $\sim 0.8 ~ - ~0.9$.

 It's to note here again that the conversion of rapidity to pseudorapidity
phase space is associated with a Jacobian  $J(y,\eta)$, which is given by the right hand side
multiplier of Eqn. \ref{eqn:eta-y-conversion}. This depends on the
momentum distribution of the produced particles. In the limit of rest
mass of the particles being much smaller than their momenta,
$J(y,\eta) =1$. The value of the Jacobian is smaller at LHC
energies, compared to that at RHIC energies, as the average transverse
momentum of particles increases with beam energy. As measured by the
PHENIX experiment, for central Au+Au collisions at $\sqrt{s_{NN}} =
200$ GeV,  $J(y,\eta) = 1.25$. Whereas, the corresponding
measurement of $J(y,\eta) = 1.09$ for Pb+Pb central collisions at
$\sqrt{s_{NN}} = 2.76$ TeV by the CMS experiment at LHC. 
Rewriting Eqn. \ref{eqn:eta-y-conversion} after taking an integration 
over $p_T$, one obtains:
\begin{equation}
\frac{dN}{d\eta} ~=~  v (y) \frac{dN}{dy}
\label{eqn:eta-y}
\end{equation}
where $v$ is the velocity of the particle.
For a hadron of mass $m$ and momentum $p$, which emerges at an angle $90^0$ with respect to 
the beam direction, $y=\eta = 0$, the above relationship becomes,
\begin{equation}
\frac{dN}{d\eta}|_{\eta = 0} ~=~  v  \frac{dN}{dy}|_{y =0}
%\label{eqn:eta-y}
\end{equation}
As most of the particles are pions in the final state, with an average momentum of three times
the pion mass, we have $v = 0.95$. At mid-rapidity, the rapidity-pseudorapidity conversion
hence involves with almost a constant factor. Hence the shape is not affected to a greater
extent. However, when one considers the whole rapidity range, where the particle velocity
in fact becomes a function of the rapidity, the shape of the pseudorapidity distribution
(which is characterized by the width) is different from the rapidity distribution.

Usually, the rapidity spectra are parametrized by the sum of two Gaussian distributions
positioned symmetrically with respect to mid-rapidity \cite{na49-onset, na49-spectra},
\begin{equation}
\boxed{\frac{dN}{dy} ~=~  \frac{\left<N\right>}{2\sqrt{2\pi \sigma^2}} \left\{ exp \bigg[-\frac{1}{2}\bigg(\frac{y-y_0}{\sigma}\bigg)^2\bigg]+
exp \bigg[-\frac{1}{2}\bigg(\frac{y+y_0}{\sigma}\bigg)^2\bigg]\right\}}
\label{eqn:eta-dist-param}
\end{equation}   
where $\left<N\right>, ~\sigma$ and $y_0$ are fit parameters. $\sigma$ is the width of the 
rapidity distribution. Landau's energy dependent Gaussian rapidity distribution is given by \cite{carruthers, wong} \\

\begin{eqnarray}
\frac{1}{\sigma_{in}} \frac{d\sigma}{dy} &=& \frac{dN}{dy} \nonumber \\
&=& \frac{N}{\left(2\pi L\right)^{1/2}}~ exp. \left(-y^2/2L\right)
\label{eqn:landaudN}
\end{eqnarray}
with parameters
\begin{eqnarray}
L&=& \frac{1}{2} ~ln\left( s/4m_p^2\right) \nonumber \\
&=& ln ~\gamma ~=~ ln \left(\sqrt{s_{NN}}/2m_p\right)
\end{eqnarray}
where $s \equiv$ squared total center of mass energy. Comparing Eqns. \ref{eqn:eta-dist-param} and \ref{eqn:landaudN}, we get \\

\begin{equation}
\boxed{\sigma_y ~=~  \sqrt{ln(\sqrt{s_{NN}}/2m_p)}}
\label{eqn:width}
\end{equation}   

\begin{figure}
\centering
\includegraphics[width=15.5cm]{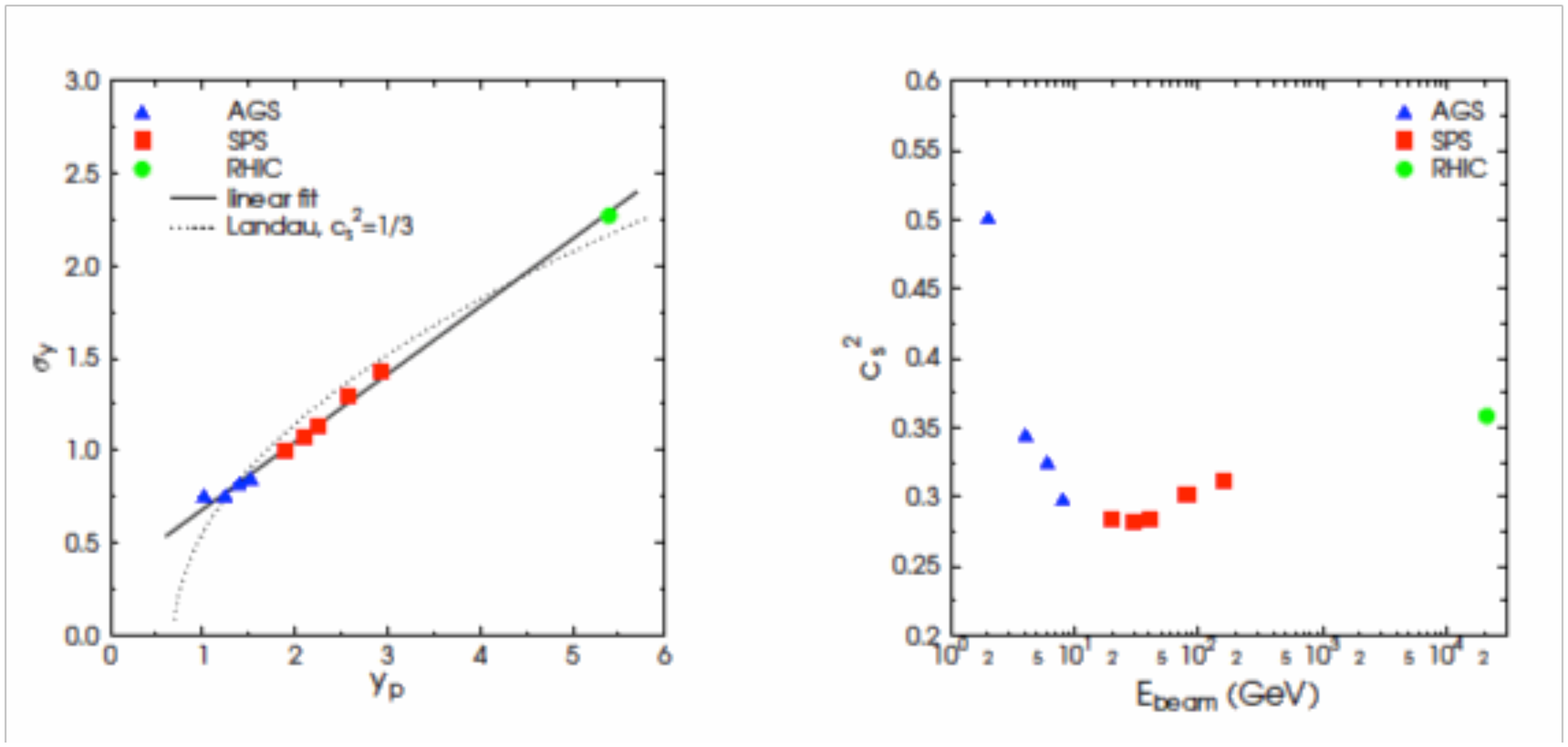}
\caption{Left: The width of rapidity distributions of $\pi^-$ in central Pb+Pb (Au+Au) collisions as a function the beam rapidity. The dotted line indicates Landau model predictions with $c_s^2 ~=~ 1/3$, while the full line shows a linear fit to data points. Right: Speed of sound as a function of beam energy showing a softest point at $E_{beam} ~=~ 30 A~GeV$ \cite{petersen}.}
\label{fig:soundVelocity}
\end{figure}

$\sigma_y$ is the width of the rapidity distribution, with $m_p$ being the mass of proton. 
Interestingly, the width of the rapidity distribution
is related to the longitudinal flow and velocity of sound in the medium and hence can 
probe the equation of state of the produced matter. However, this is affected by the
final state rescattering, which could be understood through a $p_T$ dependent study
of $\sigma_y$ to disentangle the initial hard scattering from the final state rescattering.
With the assumption that the velocity of sound $c_s$ is independent of temperature, the rapidity
density in the framework of Landau hydrodynamic model is given by \cite{soundVel,shuryak}
\begin{equation}
\frac{dN}{dy} ~=~ K ~ \frac{s_{NN}^{1/4}}{\sqrt{2\pi~\sigma_y^2}}~exp. \left(-\frac{y^2}{2\sigma_y^2}\right)
\end{equation}
where 
\begin{eqnarray}
\sigma_y^2 =\frac{8}{3}~\frac{c_s^2}{1-c_s^4}~ln\left(\sqrt{s_{NN}}/2m_p\right)  \nonumber \\ 
\Rightarrow \boxed{\sigma_y^2 = \frac{8}{3}~\frac{c_s^2}{1-c_s^4}~ln~\gamma}
\end{eqnarray}
and $K \equiv$ normalization factor. Inverting the above equation for $\sigma_y^2$, one can get \cite{petersen}\\
\begin{equation}
c_s^2 ~=~ \frac{-4}{3} ~\frac{ln\left(\sqrt{s_{NN}}/2m_p\right)}{\sigma_y^2}~+~ \sqrt{\left[\frac{4}{3}~\frac{ln\left(\sqrt{s_{NN}}/2m_p\right)}{\sigma_y^2}\right]^2~+~1}
\end{equation}
\underline{For an ideal gas (Landau model prediction) the velocity of sound is, $c_s^2 ~=~ 1/3$.} 
\underline{However, for a gas of hadrons, $c_s^2 ~=~ 1/5$.} This implies, the expansion of the system is slower compared to an ideal gas scenario. The equation of state is given by\\

\begin{equation}
\boxed{\frac{\partial P}{\partial \epsilon}~=~ c_s^2}
\end{equation}
where, $P$ is the pressure and $\epsilon$ is the energy of the system under consideration. It should be noted here that when the expansion of the matter proceeds as longitudinal and superimposed transverse expansions, a rarefaction wave moves radially inwards with the velocity of sound. The velocity of sound in the medium formed in heavy-ion collisions, when studied as a function of collision (beam) energy, shows the softest point occurring around $E_{beam}~=~ 30$ AGeV. This could be a signature of deconfinement transition \cite{petersen}.

\begin{center}
\begin{tabular}{l|l}
\hline
\multicolumn{2}{c}{\bf{Notation} \cite{bhalerao}} \\
\cline{1-2}
\multicolumn{1}{c|}{Coordinate Space} & \multicolumn{1}{c}{Momentum Space}\\
\hline
$(t,x,y,z)$ $\rightarrow$  ($\tau, r, \phi, \eta$)&  $(E, p_x, p_y,
p_z)$ $\rightarrow$ ($m_T, \vec{p_T}, y$) \\[4mm]
$\tau = \sqrt{t^2-z^2}$ = proper time & $m_T = \sqrt{E^2-p_z^2}$ =
transverse mass\\[4mm]
$\eta = tanh^{-1}\frac{z}{t}$ = space-time rapidity &
$y=tanh^{-1}\frac{p_z}{E}=tanh^{-1}v_z$ = rapidity \\ [4mm]
Conversely, & Conversely, \\ [4mm]
$t=\tau ~ cosh~\eta,$ $z=\tau~ sinh~\eta$ & $E=m_T ~cosh~y,$ $p_z=m_T~
sinh~y$\\ [4mm]

\hline
\end{tabular}
\end{center}
 
\subsection{The Invariant Yield}
The rapidity variable has the useful property that it transforms linearly under a Lorentz transformation so that the invariant differential single particle inclusive cross section becomes:

\begin{equation}
\boxed{\frac{Ed^3\sigma}{dp^3} ~=~ \frac{Ed^3\sigma}{p_Tdp_Tdp_Ld\phi} ~=~\frac{d^3\sigma}{p_Tdp_Tdyd\phi}},
\end{equation}
where
\begin{equation}
dy~=~ \frac{dp_L}{E}
\end{equation}

First we proceed to show $\frac{d^3p}{E}$ is Lorentz invariant. The differential 
of Lorentz boost in longitudinal direction is given by
\begin{equation}
dp_z^* ~=~\gamma(dp_z-\beta dE).
\label{dp}
\end{equation}
Taking the derivative of the equation $E^2 ~=~ p^2 + m^2$, we get
\begin{equation}
EdE ~=~p_zdp_z.
\label{de}
\end{equation}
Using equations \ref{dp} and \ref{de} we get
\begin{eqnarray}
dp_z^* &=& \gamma(dp_z -\beta \frac{p_zdp_z}{E}) \nonumber \\
&=& \frac{dp_z}{E}~E^* .
\end{eqnarray}
As ${\bf p}_T$ is Lorentz invariant, multiplying ${\bf p}_T$ on both the sides 
and re-arranging gives
\begin{equation}
\frac{d^3p^*}{E^*} ~=~\frac{d^2 {\bf p}_T~dp_z}{E} ~=~ \frac{d^3p}{E}.
\end{equation}
In terms of experimentally measurable quantities, $\frac{d^3p}{E}$ could be 
expressed as
\begin{eqnarray}
\frac{d^3p}{E} &=& d{\bf p}_T~dy \nonumber \\
&=& p_Tdp_Td\phi dy \\
&=& m_Tdm_Td\phi dy .
\end{eqnarray}
The Lorentz invariant differential cross-section
$\frac{Ed^3\sigma}{dp^3} ~=~ \frac{Ed^3N}{dp^3}$ is the invariant yield. In 
terms of experimentally measurable quantities this could be expressed as
\begin{eqnarray}
\frac{Ed^3\sigma}{dp^3} &=& \frac{1}{m_T} ~\frac{d^3N}{dm_Td\phi dy} 
\nonumber \\
&=&  \frac{1}{2\pi~m_T} ~\frac{d^2N}{dm_T dy} \nonumber \\
&=&  \frac{1}{2\pi~p_T} ~\frac{d^2N}{dp_T dy}  .
\label{inv}
\end{eqnarray}
To measure the invariant yields of identified particles equation \ref{inv} is 
used experimentally.

\subsection{Inclusive Production of Particles and the Feynman Scaling 
variable $x_F$}
A reaction of type\\
$beam ~+~ target ~ \longrightarrow ~ A ~+~ anything$\\
where $A$ is known is called an ``{\it inclusive reaction}''. The cross-section 
for particle production could be written separately as functions of ${\bf p}_T$ 
and $p_L$:\\
\begin{equation}
\sigma ~=~f({\bf p}_T) g(p_L).
\end{equation}
This factorization is empirical and convenient because each of these factors 
has simple parametrizations which fit well to experimental data.\\

Similarly the differential cross-section could be expressed by
\begin{equation}
\frac{d^3\sigma}{dp^3} ~=~ \frac{d^2\sigma}{{\bf p}_T^2}~ \frac{d\sigma}{dp_L}
\end{equation}
Define the variable 
\begin{eqnarray}
x_F = \frac{p_L^*}{p_L^*(max)}  \nonumber \\
\Rightarrow \boxed{x_F = \frac{2p_L^*}{\sqrt{s}}}
\end{eqnarray}
$x_F$ is called the {\it Feynman scaling variable}: longitudinal component of 
the cross-section when measured in CMS of the collision, would scale {\it i.e.} 
would not depend on the energy $\sqrt{s}$. This is the fraction of
maximum allowed longitudinal momentum ($-1 \leq x_F \leq$ 1) carried
by the particle in the CMS. This is used in comparing the shapes of particle distributions at different collision energies near the projectile or target rapidity. Instead of $\frac{d\sigma}{dp_L^*}$, $\frac{d\sigma}{dx_F}$ 
is measured which wouldn't depend on energy of the reaction, $\sqrt{s}$. This 
Feynman's assumption is valid approximately. This variable is usually
used to compare particle distribution at different collision energies.\\

  The differential cross-section for the inclusive production of a particle is 
then written as 
\begin{equation}
\frac{d^3\sigma}{dx_Fd^2{\bf p}_T} ~=~ F(s,~x_F,~{\bf p}_T)
\end{equation}
Feynman's assumption that at high energies the function $F(s,~x_F,~{\bf p}_T)$ 
becomes asymptotically independent of the energy means:\\

$lim_{s \rightarrow \infty} F(s,~x_F,~{\bf p}_T)~=~ F(x_F, ~{\bf p}_T)~=~ 
f({\bf p}_T)~g(x_F)$

According to Feynman, the mean number of particles rises logarithmically in the asymptotic limit of large energies. This in fact applies to any kind of particles:\\

$<N> ~\propto ln~W ~\propto ln~\sqrt{s}$,\\

 where $W= \sqrt{s}/2$, is the beam energy of a symmetric collider. 
However, these conclusions are based on phenomenological arguments about the exchange of quantum numbers between the colliding particles. It was assumed that the number of particles with a given mass and transverse momentum ($\bf {p}_{\rm T}$) in a longitudinal interval $p_z$ depends on the energy\\
$E ~ = ~ E(p_Z)$ as
\begin{equation}
\frac{dN}{dp_z} \sim \frac{1}{E}
\end{equation}

This means the among all the produced particles, the probability of finding a particle of kind $i$ with transverse momentum $\bf {p}_{\rm T}$, mass $m$ and longitudinal momentum $p_z$ is of the form:\\

\begin{equation}
f_i ({\bf p}_{\rm T}, ~x_F=p_z/W) ~\frac{dp_z}{E}~ d^2p_T
\label{feynmanDist}
\end{equation}
with $E~=~\sqrt{m^2+p_T^2+p_z^2}$ being the total energy of the particle under discussion. The function $f_i({\bf p}_{\rm T}, ~x_F)$ denotes the particle distribution.  Let's derive Eqn. \ref{feynmanDist} to have a hand on experience.\\

Rewriting Eqn. \ref{feynmanDist} in the form of invariant cross section
\begin{equation}
\frac{1}{\sigma} ~ E \frac{d^3\sigma} {dp_z~d^2{\bf p}_T} ~ =~ f_i({\bf p}_{\rm T}, ~x_F) 
\label{feynmanDist1}
\end{equation}
$f_i$ factorizes approximately (found experimentally) and a normalization $g_i$ is chosen such that

\begin{eqnarray}
\int f_i ({\bf p}_{\rm T}, ~x_F)~d^2p_T  &=&  f_i(x_F)\int g_i({\bf p}_T)~ d^2p_T \nonumber \\
&=& f(x_F)
\label{feynmanDist2}
\end{eqnarray}
Here 
\begin{equation}
\int g_i({\bf p}_T)~ d^2p_T ~=~ 1
\end{equation}
Now, integrating Eqn. \ref{feynmanDist1}  and applying Eqn. \ref{feynmanDist2}, we get
\begin{eqnarray}
\int \frac{1} {\sigma} ~ E ~ \frac{d^3 \sigma} {dp_z~d^2{\bf p}_T} ~ \frac{d^3p} {E} &=& <N> \nonumber \\
&=&  \int f_i ({\bf p}_{\rm T}, ~x_F)~\frac{d^3p}{E} \nonumber \\
&=&  \int f_i(x_F) \frac{dp_z}{\sqrt{W^2x_F^2 + m_T^2}}
\label{feynmanDist3}
\end{eqnarray}
On the left hand side we have used the definition of invariant cross section with the average particle multiplicity $<N>$, and for $m_{\rm T}$ an effective average $p_{\rm T}$ is used with $x_{\rm F} ~=~ p_{\rm z}/W$. Hence
\begin{equation}
<N>~ =~ \int_{-1}^{+1} \! f_i(x_{\rm F}) \, \frac{\mathrm{d}x_F}{\sqrt{x_F^2 + \frac{m_T^2}{W^2}}},     
\label{feynmanDist4}
\end{equation}
with $dx_{\rm F} ~=~ \frac{dp_{\rm z}}{W}$.\\

The integral is symmetric because $f_i(x_{\rm F})$ is symmetric for collisions of identified particles. For asymmetric collision systems, the integration could be performed separately for negative and positive $x_{\rm F}$ and yields the same result.\\

$f_i (x_{\rm F}) \leq ~ B$ is finite and bounded due to energy conservation. Feynman assumed that for $x_{\rm F} ~=~ 0$ a finite limit is reached.\\
\begin{eqnarray}
x_{\rm F} \rightarrow 0 ~~ \Rightarrow ~~ \frac{p_z}{W} &=& \frac{p_z}{E_{lab}} ~~\rightarrow ~~ 0 \nonumber \\
&\Rightarrow& E_{lab} \rightarrow \infty \nonumber
\end{eqnarray}

Hence

\begin{eqnarray}
<N> &=& 2 ~ \int_{0}^{1} \! f_i(x_{\rm F}) \, \frac{\mathrm{d}x_F}{\sqrt{x_F^2 + \frac{m_T^2}{W^2}}} ~ 
\leq ~ 2 ~ \int_{0}^{1} \! B \, \frac{\mathrm{d}x_F}{\sqrt{x_F^2 + \frac{m_T^2}{W^2}}} \nonumber \\
&=& 2B ~ ln \left[ x_F + \sqrt{x_F^2 + \frac{m_T^2}{W^2}}\right]\Big|_0^1 \nonumber \\
&=& 2B ~ ln \left[ 1 + \sqrt{1 + \frac{m_T^2}{W^2}}\right] - 2B~ ln \left(\frac{m_T}{W} \right)
\end{eqnarray}
In the limit $W \rightarrow \infty$, the first term of the above equation could be shown to be constant and the second term is proportional to $ln~W$. hence, Feynman scaling tells that the average total multiplicity scales as:\\

\begin{equation}
\boxed{<N> ~ \propto ~ ln ~W ~ \propto ~ ln~\sqrt{s}}
\end{equation}

If one considers the maximum reachable rapidity in a collision to increase like $ln ~\sqrt{s}$ ($y_{\rm {max}}~ \sim ~ ln ~ \sqrt{s}$) and in addition, particles are evenly distributed in rapidity, it follows that $\frac{dN}{dy}$ is independent of $\sqrt{s}$:\\

\begin{equation}
\boxed{\frac{dN}{dy} ~ =~ Constant}
\end{equation}

{\bf Feynman's Assumption:} $f_i(p_{\rm T}, x_{\rm F})$ which denotes the particle distribution, becomes independent of $W$ at high energies. This assumption is known as \underline{Feynman Scaling} and $f_i$ is called the scaling function or Feynman function. The variable\\

\begin{equation}
\boxed{x_F ~ = p_z/W ~=~ 2p_z/\sqrt{s}}
\end{equation}
is called \underline{\it Feynman-x}.\\

{\it Feynman-x} is the ratio of the longitudinal momentum of the particle to the total energy of the incident particle.
{\bf Note:} \\

1. $\frac{dN}{dy} ~ =$ Constant $\Rightarrow$ Height of the rapidity distribution around mid-rapidity (the so-called plateau) is independent of collision energy, $\sqrt{s}$. \\

2. Equivalently, if  Feynman scaling holds good,  the pseudorapidity density of charged particles at mid-rapidity {\it i.e.} $\frac{dN}{d\eta} (\eta = 0)$ is approximately constant.\\

 Scaling properties of various distributions can also be studied in terms of the scaled rapidity:\\
 \begin{equation}
\boxed{z= y^*/y^*_{\rm max}.}
\end{equation}
 
 The two scaled variables: $x_{\rm F}$ and $z$ emphasize different kinematic regions: the detailed structure of the central part of the distribution ({\it i.e.} large emission angles) can be better seen using $x_{\rm F}$, while the far {\it "wings"} ({\it i.e.} small angles) using $z$.

\subsection{What is a ${\bf p}_{\rm T}$ or $ m_{\rm T}$ Spectrum?}
${\bf p}_{\rm T}$ spectrum: $\frac{1}{2\pi p_T} \frac{d^2N}{dy~dp_T} $ Vs $p_{\rm T}$ \\

$m_{\rm T}$ spectrum: $\frac{1}{2\pi m_T} \frac{d^2N}{dy~dm_T} $ Vs $m_{\rm T}$ or ($m_{\rm T}-m_0$) \\

The invariant yield when plotted as a function of ${\bf p}_{\rm T}$ or $ m_{\rm T}$ or ($m_{\rm T}-m_0$) is called ${\bf p}_{\rm T}$ or $ m_{\rm T}$ spectrum, respectively. Experimentally at kinetic freeze-out when the elastic collisions between the final state particles almost cease to happen (in other words the particle mean free path becomes higher than the system size: size of the produced fireball) then the ${\bf p}_{\rm T}$ or $ m_{\rm T}$-spectrum is frozen, which carries the kinetic freeze-out properties of the system. Recall here that $\boxed{p_{\rm T}dp_{\rm T} ~=~ m_{\rm T}dm_{\rm T}}$.  

The distribution of particles as a function of ${\bf p}_T$ is called 
${\bf p}_T$-distribution.
Mathematically, 
\begin{equation}
\frac{dN}{d{\bf p}_T} ~=~ \frac{dN}{2\pi~|{\bf p}_T|d|{\bf p}_T|}
\end{equation}
where $dN$ is the number of particles in a particular ${\bf p}_T$-bin. People 
usually plot $\frac{dN}{p_Tdp_T}$ as a function of $p_T$ taking out the factor 
$1/2\pi$ which is a constant. Here $p_T$ is a scalar quantity. The low-$p_T$ 
part of the $p_T$-spectrum is well described by an exponential function having 
thermal origin given by Eqn. (\ref{expon}). However, a QCD-inspired
power-law function (given by Eqn. (\ref{powerlaw})) seems to provide a
better description of the high $p_T$ ($\gtrsim 3$ GeV/c) region. To
describe the whole range of the $p_T$-spectrum, one uses the Levy
function given by Eqn.(\ref{levy}) which has an exponential part to describe low-$p_T$ and a 
power-law function to describe the hight-$p_T$ part which is dominated by hard 
scatterings (high momentum transfer at early times of the collision). 
\begin{equation}
\frac{1}{2\pi p_T} \frac{d^2N}{dy~dp_T} ~=~ A~e^{\frac{-m_T}{T}},
\label{expon}
\end{equation}

\begin{equation}
\frac{1}{2\pi p_T} \frac{d^2N}{dy~dp_T} ~=~ B\left (1+\frac{p_T}{p_0}\right)^{-n},
\label{powerlaw}
\end{equation}

\begin{equation}
\frac{1}{2\pi p_T} \frac{d^2N}{dy~dp_T} ~=~
\frac{dN}{dy}\frac{(n-1)(n-2)}{2\pi nC[nC+m_0(n-2)]} \times \left( 1+
 \frac{\sqrt{p_T^2 + m_0^2}-m_0}{nC} \right)^{-n},
\label{levy}
\end{equation}
where $A, T, B, p_0, n, \frac{dN}{dy}, C,$ and $m_0$ are fit
parameters \cite{starStrange}.
The inverse slope parameter of $p_T$-spectra is called the effective temperature 
($T_{eff}$), which has a thermal contribution because of the random kinetic 
motion of the produced particles and a contribution from the collective motion 
of the particles. This will be described in details in the section of 
freeze-out properties and how to determine the chemical and kinetic freeze-out 
temperatures experimentally.\\

If we look deeper into Eqn. \ref{expon}, we expect this for a 2-dimensional classical 
thermalized fluid at rest. We recall here that Boltzmann distribution $\propto ~exp. (-\beta E)$, where $\beta ~\equiv ~ 1/T$ is the inverse temperature. We use a semi-logarithmic plot ($y$-axis logarithmic scale) of $N~ exp. \left(-\beta m_T\right)$ vs $(m_T-m_0)$ which happens to be a straight line with slope $-\beta$, from which the effective temperature is extracted. A semi-log plot of $N~ exp. \left(-\beta m_T\right)$ vs $p_T$ is an approximate straight line if $p_T ~\gg~ m_0$. 

%$$$$$$$$$$$$$$$$$$$$$$$$$$$$$$$$$$$$$$$$$$$$$$$$$$$$$$$$$$$$$$$$$$$$

The most important parameter is then the mean $p_T$ which carries the 
information of the effective temperature of the system. Experimentally, 
$\left<p_T\right>$ is studied as a function of $\frac{dN_{ch}}{d\eta}$ which 
is the measure of the entropy density of the system. This is like studying the 
temperature as a function of entropy to see the signal of phase transition. 
The phase transition is of 1st order if a plateau is observed in the spectrum 
signalling the existence of latent heat of the system (like
liquid-vapour phase transition). This was first proposed
by L. Van Hove \cite{vanHove}.\\

The average of any quantity $A$ following a particular probability distribution 
$f(A)$ can be written as
\begin{equation}
\left<A\right> ~=~ \frac{\int A~f(A)~dA}{\int f(A)~dA}.
\end{equation}
Similarly,
\begin{eqnarray}
\left<p_T\right> &=& \frac{\int_0^{\infty} p_T~(\frac{dN}{dp_T})~dp_T}
{\int_0^{\infty} (\frac{dN}{dp_T})~dp_T} \nonumber \\ 
&=&  \frac{\int_0^{\infty} p_T~dp_T~~p_T(\frac{dN}{p_Tdp_T})}
{\int_0^{\infty}p_T~dp_T (\frac{dN}{p_Tdp_T})} \nonumber \\ 
&=&  \frac{\int_0^{\infty} p_T~dp_T~~p_T~f(p_T)}{\int_0^{\infty}p_T~dp_T f(p_T)}
\end{eqnarray}
where $2\pi~p_T~dp_T$ is the phase space factor and the $p_T$-distribution 
function is given by 
\begin{equation}
f(p_T) ~=~ \frac{dN}{d{\bf p}_T} ~=~ \frac{dN}{p_Tdp_T} .
\end{equation}
%\hline
\underline{{\bf Example}}
Experimental data on $p_T$-spectra are sometimes fitted to the exponential 
Boltzmann type function given by
\begin{equation}
f(p_T) ~=~ \frac{1}{p_T}\frac{dN}{dp_T} ~\simeq~ C~ e^{-m_T/T_{eff}}.
\end{equation}
%\hline

The $\left<m_T\right>$ could be obtained by
\begin{eqnarray}
\left<m_T\right> &=& \frac{\int_0^{\infty}p_T ~dp_T~m_T~exp.(-m_T/T_{eff})}
{\int_0^{\infty}p_T ~dp_T~exp.(-m_T/T_{eff})} \nonumber \\
&=& \frac{2T_{eff}^2+2m_0T_{eff}+m_0^2}{m_0+T_{eff}}\\
&\Rightarrow& \boxed{\left<m_T\right> = T_{eff}+m_0 + \frac{(T_{eff})^2}{m_0+T_{eff}}}
\end{eqnarray} 
where $m_0$ is the rest mass of the particle. It can be seen from the above 
expression that for a massless particle
 \begin{equation}
\left<m_T\right> ~=~ \left<p_T\right>  ~=~ 2T_{eff} .
\end{equation}
This also satisfies the principle of equipartition of energy which is expected 
for a massless Boltzmann gas in equilibrium.  However, in experiments the 
lower (higher) limit of $p_T$ is a finite quantity. In  that case the integration will 
involve an incomplete gamma function.

\subsection{How is the radial flow measured from $p_{\rm T}$-spectra?}
In central heavy-ion collisions, radial flow is supposed to play a vital role in the thermodynamic expansion of the produced fireball. Radial flow is related to the initial pressure produced just after the collision. This could be extracted from the analysis of the transverse momentum spectra. Assuming a thermalized non-relativistic plasma (for simplicity),  particle velocity \cite{bhalerao}

\begin{equation}
\vec{v} ~=~ \vec{v}_{flow} ~+~ \vec{v}_{th},
\end{equation}

where $\vec{v}_{flow} ~\equiv ~$ transverse velocity of the expanding fluid, which is independent of the particle species and is the collective component of $\vec{v}$. \\

$\vec{v}_{flow} ~\equiv ~$ thermal component of $\vec{v}$, which is generated due to random thermal motion of the quanta of the system.\\

Hence for a particle of mass $m_0$\\
\begin{eqnarray}
\left\langle  \frac{1}{2}~m_0 v^2\right\rangle &=& \frac{1}{2}~m_0 v_{flow}^2 ~+~ \left\langle \frac{1}{2}~m_0 v_{th}^2\right\rangle \nonumber \\
&=& \frac{1}{2}~m_0 v_{flow}^2 ~+~  \frac{3}{2} ~kT,
\end{eqnarray}
where $T$ is the temperature of the fluid. Hence the average kinetic energy (K.E) is given by 

\begin{eqnarray}
\left\langle  K.E.\right\rangle  &=& \frac{1}{2}~m_0 v_{flow}^2 ~+~  \frac{3}{2} ~kT \nonumber \\
\Rightarrow \frac{3}{2} ~kT_{eff} &=& \frac{1}{2}~m_0 v_{flow}^2 ~+~  \frac{3}{2} ~kT_{th} \nonumber \\
\Rightarrow T_{eff} &=& T_{th} ~+~ \frac{1}{3}~m_0 v_{flow}^2  ~(taking ~k=1) 
\end{eqnarray}

Because the final state particle are measured at freeze out (after they stream out to reach the detectors), the extracted values of $\vec{v}_{flow}$ and $T$ correspond to the instant of freeze out.\\

Taking $k=1$, in 2-dimension: \\
\begin{equation}
 \boxed{T_{eff} = T_{th} ~+~ \frac{1}{2}~m_0 v_{flow}^2},
\end{equation}

and in 1-dimension:\\
\begin{equation}
 \boxed{T_{eff} = T_{th} ~+~ m_0 v_{flow}^2}
\end{equation}
Note that these formulae are used essentially when the spectra is well described by an exponential function
(low-$p_{\rm T}$ regime). However, when one goes to high-$p_{\rm T}$ regime, the following formula could be used for extracting the radial flow from the $p_{\rm T}$ or $m_{\rm T}$-spectra.\\

\begin{equation}
 \boxed{T_{eff} = T_{th} ~\sqrt{\frac{1+v_{flow}}{1-v_{flow}}}}
\end{equation}

\vspace{3cm} 

\vskip-0.75cm
% Create the reference section using BibTeX:

\end{document}